\documentclass[letterpaper]{article}

\usepackage[T1]{fontenc}
\usepackage{geometry}
\usepackage{setspace}

\usepackage{achemso}
\usepackage{natbib}

\usepackage{graphicx}
\usepackage{float}
\usepackage{url}
\newfloat{scheme}{htbp}{los}
\floatname{scheme}{Scheme}
\floatname{chart}{Chart}
\newfloat{graph}{htbp}{loh}

\usepackage{placeins}
\usepackage{textcomp}
\usepackage{textgreek}

\usepackage{chemformula} % Formulas using \ch{}
\usepackage[version = 4]{mhchem} % Formulas using \ce{}
\usepackage{siunitx}
\usepackage{braket}
\usepackage{multirow}

\usepackage{tikz}
\usetikzlibrary{decorations.pathreplacing,calc,positioning}
\usepackage{authblk}
\author[1]{David David$^{\$*}$}
\affil[1]{Capgemini Quantum Lab}
\author[2]{Vedangi Pathak$^{\$\dagger}$}
\affil[2]{IBM T. J. Watson Research Center, Yorktown Heights, NY}
\author[1]{Marek Kowalik$^\$$}
\author[2]{Hamed Mohammadbagherpoor}
\author[2]{Vincent Beltrani}
\author[3]{Kara Maller}
\affil[3]{Q-CTRL}
\author[3]{Niall Moroney}
\author[1]{Phalgun Lolur$^\ddagger$}

\title{Sample-Based Quantum Diagonalization for Open-Shell Transition-Metal Complexes in Gas and Implicit Solvent}
\date{*Email: david.david@capgemini.com \\ $\dagger$Email: vedangi.pathak@ibm.com \\ $\ddagger$ Email: phalgun.lolur@capgemini.com}

\begin{document}

\maketitle

\begin{abstract}

Open-shell $3d$ transition-metal complexes pose a stringent challenge for correlated electronic-structure methods because near-degenerate metal $d$ orbitals give rise to competing spin states and oxidation-state-dependent charge transfer, while solvation can reshape the relative stability of the resulting electronic configurations. Building on prior demonstrations of open-shell sample-based quantum diagonalization (SQD) for light-atom systems and, separately, closed-shell SQD coupled to the integral-equation-formalism polarizable continuum model (IEF-PCM), we combine these capabilities through an open-shell SQD--IEF-PCM framework and apply it to the octahedrally coordinated $\mathrm{[Co(H_2O)_5CO_2]^{2+/3+}}$ complex, spanning the Co(III) singlet and quintet and Co(II) doublet and quartet states along a metal--ligand dissociation coordinate. Using quantum samples collected on an IBM Heron processor and active spaces of up to 50 qubits, SQD closely reproduces coupled-cluster and heat-bath configuration-interaction benchmarks within the same active spaces in both gas phase and implicit solvent; for high-spin quintet $\mathrm{[Co(H_2O)_5CO_2]^{3+}}$, it additionally resolves an avoided crossing associated with internal charge transfer that is absent in the corresponding singlet and in the lower oxidation state, while IEF-PCM suppresses this crossover by preferentially stabilizing the dissociation channel leading to neutral CO$_2$. To our knowledge, this work provides the first hardware demonstration of SQD for an open-shell 3d transition-metal complex with discrete metal-centered d-orbital chemistry, and the first application of open-shell SQD in a dielectric continuum, establishing a quantum-centric framework for transition-metal chemistry in regimes where spin-state energetics, charge transfer, and environmental response are strongly coupled.
\end{abstract}

\section*{Keywords}

Sample-based quantum diagonalization; open-shell electronic structure; transition-metal complexes; charge transfer; implicit solvation; polarizable continuum model; quantum computing, quantum selected configuration interaction

\footnote{$^\$$These authors contributed equally to this work.}

%%%%%%%%%%%%%%%%%%%%%%%%%%%%%%%%%%%%%%%%%%%%%%%%%%%%%%%%%%%%%%%%%%%%%
%% Start the main part of the manuscript here.
%%%%%%%%%%%%%%%%%%%%%%%%%%%%%%%%%%%%%%%%%%%%%%%%%%%%%%%%%%%%%%%%%%%%%

\begin{figure}[H]
    \centering
    \includegraphics[width=1\linewidth]{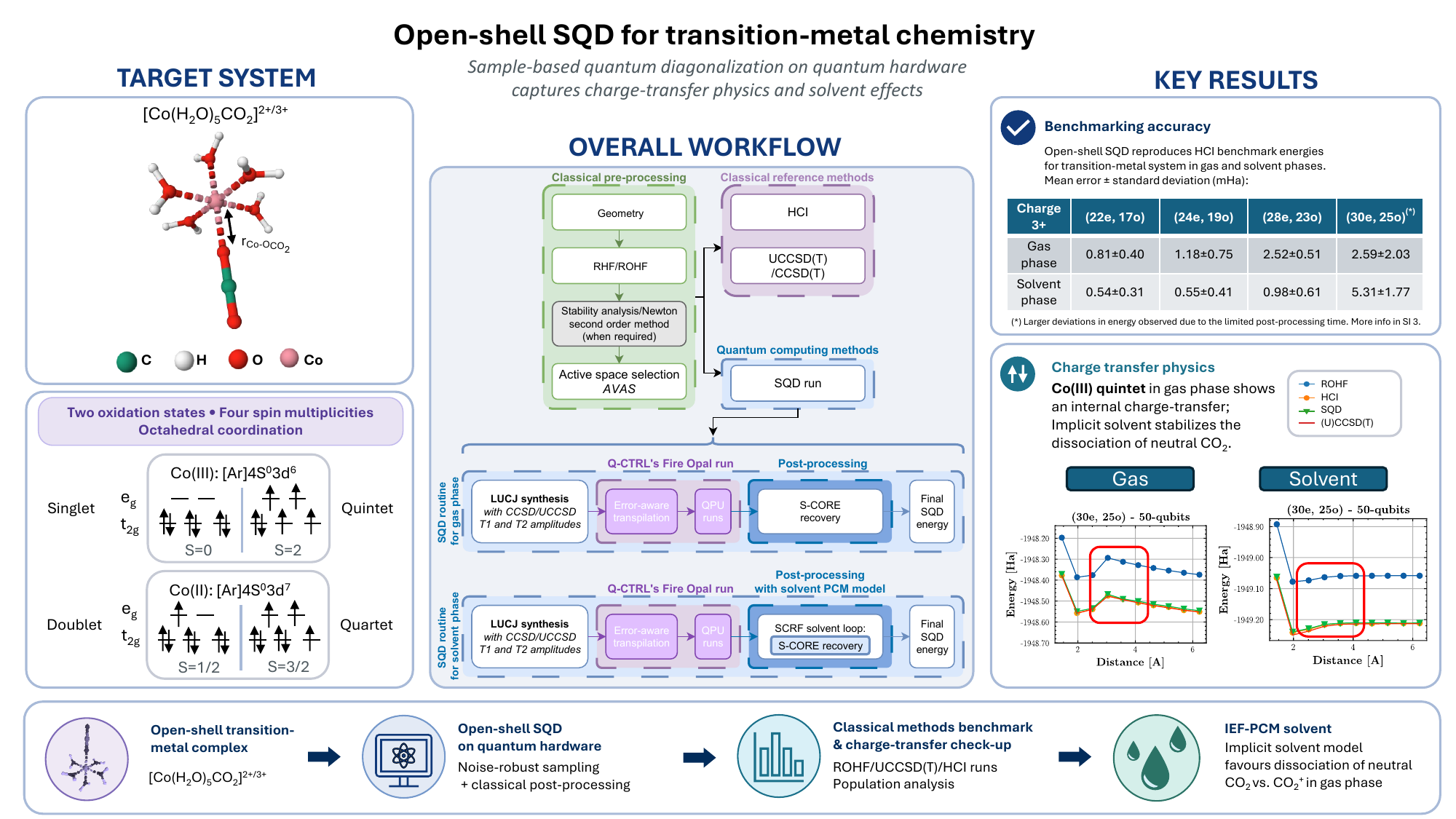}
    \caption{\textbf{Overview of the open-shell SQD study of a transition-metal complex}. The $\mathrm{[Co(H_2O)_5CO_2]^{2+/3+}}$ test system, spanning two oxidation states and four spin manifolds is presented in the left panel, as a representative platform for investigating open-shell transition-metal chemistry. Render shows the relaxed geometry. The  $\mathrm{Co-O_{(CO_2)}}$ distance is indicated, being the distance parameter of the dissociation curve. 
    Middle section of the figure summarizes the overall workflow combining open-shell SQD on IBM quantum hardware with classical electronic-structure benchmarks and implicit-solvent modeling. 
    For classical pre-processing the first step is geometry optimization. Second step performs RHF for singlet spin state, otherwise it performs ROHF. Then stability-analysis and second-order SCF is carried out where needed and finally an active space is constructed using the AVAS method.
    Classical reference methods (HCI, UCCSD(T)/CCSD(T)) are evaluated in parallel for benchmarking. Lower part of the workflow shows details of the SQD protocol for gas phase and the implicit solvent, separately. For gas-phase standard SQD self-consistent recovery post-processing iteratively yields the adjusted subspace, ground-state and its energy. For solvent phase, SQD iterations construct an effective subspace while self-consistently updating the solute-solvent interaction potential within a PCM model, yielding the final SQD energy. 
    Section on the right captures the key results of this work. Upper panel highlights the mean error and standard deviation between the SQD and HCI energies for different active space sizes and multiplicities along the reaction coordinate for $\mathrm{[Co(H_2O)_5CO_2]^{3+}}$. Lower panel illustrates the charge-transfer physics observed.}
    
    \label{fig:main_figure}
\end{figure}

\section{Introduction} 

Transition-metal complexes pose demanding tests for electronic-structure methods because their properties and reactivity often depend on subtle energy differences between competing spin states, on the balanced treatment of electron correlation and metal-ligand charge-transfer character, and on environment-dependent stabilization effects arising from solvation and other condensed-phase interactions~\cite{radon2023benchmarks, reimann2022spin, may2024new, drosou2022reconciling}.
% Transition-metal complexes pose a demanding test for any electronic-structure method because their chemistry is often governed by small differences between competing spin states, by metal-ligand charge-transfer configurations, and by environment-dependent stabilization of localized charge. 
These features place transition-metal chemistry at the heart of catalysis, energy conversion, bioinorganic function, and functional materials~\cite{Sahoo2024, AbuRahman2023, James2023}, and they are also what makes the underlying wave functions challenging to compute. Even a century after the formulation of quantum mechanics, finding the lowest eigenvalue of the molecular electronic Hamiltonian under the Born-Oppenheimer approximation remains an outstanding challenge for transition-metal systems.

In the case of transition metals, the challenge is not the size of the Hilbert space alone, but its qualitative structure. Near-degenerate metal $d$ orbitals, partially filled shells or bond-breaking lead to static correlation that is usually very challenging for a single Slater determinant (SD) to describe~\cite{Sayfutyarova2017}. Hartree-Fock (HF) and Density Functional Theory (DFT) treat electron-electron interactions through a mean-field approximation. For the description of dynamic electron correlation post-HF methods such as configuration interaction (CI) and coupled cluster (CC) truncated at single and double excitations (CISD, CCSD) are usually used. However for the treatment of the static electron correlation, whenever the description of the electronic structure using a single SD is not adequate, multi-reference methods such as Complete Active-Space Self-Consistent Field (CASSCF) are used. In CASSCF CI and orbital coefficients are optimized simultaneously and its application with conventional implementation is usually limited to small active space sizes, e.g. CAS(22e,22o)~\cite{Vogiatzis2017}, due to unfavorable scaling with CAS size.  Full CI, the exact solution within a basis set, scales combinatorially as $\binom{M_o}{N_e}$, where $N_e$ is the number of electrons and $M_o$ the number of orbitals available for occupation, making it impractical for large systems. However, Hamiltonian matrix elements between determinants span many orders of magnitude, from approximately $100$~mHa to $1$~$\mu$Ha~\cite{Holmes2016}. This motivates selected CI (SCI) methods~\cite{Huron1973, Holmes2016, Tubman2020}, such as Heat-Bath CI (HCI)~\cite{Holmes2016}, which construct a compact variational determinant space rather than including all possible excitations within the chosen active space.

In theory the problem of accurately solving the Schrödinger equation is a promising task for a quantum computer \cite{Feynman1982}. The exponentially growing Hilbert space with system size could natively be represented by the quantum mechanical states of the qubits. However, in the current era of Noisy Intermediate-Scale Quantum (NISQ) devices exact algorithms like Quantum Phase Estimation (QPE)\cite{Abrams1999} are limited in their application to very small systems like H$_2$\cite{Yamamoto2025}. On the other hand, hybrid quantum-classical variational algorithms such as the Variational Quantum Eigensolver (VQE) \cite{Peruzzo2014} suffer from deep circuits and measurement overhead, especially when a Unitary CCSD \cite{Romero2018} or another chemically meaningful ansatz that yields many parameters to adjust is used.

SQD~\cite{RobledoMoreno2025} builds on the quantum-selected configuration-interaction (QSCI)~\cite{Kanno2023} strategy, in which bitstrings sampled from a quantum-prepared state are used to select determinants defining a compact configuration-interaction subspace. In practical hardware implementations, SQD combines this sampling strategy with a hardware-motivated ansatz, such as the local unitary cluster Jastrow (LUCJ) ansatz~\cite{Motta2023}, and a self-consistent configuration-recovery procedure to mitigate errors in the sampled bitstrings before the molecular Hamiltonian is diagonalized classically within the selected subspace~\cite{RobledoMoreno2025}. 
Hardware-induced violations of total particle number and $\hat S_z$ are corrected through a self-consistent configuration recovery loop, which makes SQD substantially more robust to noise than VQE and far less depth-demanding than QPE. In this study, we focus on a particularly challenging regime for quantum electronic structure methods: open-shell transition-metal chemistry in which spin state ordering, internal charge transfer, and environmental response are simultaneously active. We demonstrate that SQD can be extended to address this coupled problem within a single, consistent framework.

Recent SQD demonstrations have established the method across several distinct chemical regimes, targeting one facet of transition-metal chemistry at a time. Iron-sulfur clusters of up to $(54e,36o)$ and the dissociation of \ce{N2} have shown that SQD reaches active spaces well beyond exact diagonalization on current Heron processors~\cite{RobledoMoreno2025}. The methylene singlet-triplet gap provided the first open-shell SQD demonstration and revealed that open-shell SQD is meaningfully harder than the closed-shell case. When $N_\alpha \neq N_\beta$, the sampled determinants are no longer closed under spin inversion, the symmetrization that pins $\langle \hat S_z \rangle = 0$ in closed-shell SQD must be disabled, and the resulting subspace is no longer automatically an eigenspace of $\hat S^2$~\cite{Liepuoniute2025}. The benchmark itself was on a light-atom carbene rather than a transition-metal $d$-orbital problem. Closed-shell SQD has been embedded in an IEF-PCM self-consistent reaction-field workflow~\cite{Kaliakin2025}, and supramolecular and hydrogen-abstraction studies have probed dispersion- and radical-bond-breaking regimes~\cite{Kaliakin2025intermolecular, Smith2025}. What has not yet been attempted is the combination - open-shell metal $d$-orbital physics, an internal charge-transfer process, and an environment-dependent stabilization of charged states - within a single transition-metal problem. That combination is closer to the regime in which transition-metal chemistry is actually performed. While prior quantum-computing studies have demonstrated transition-state and bond-breaking chemistry, these have largely focused on closed-shell or light-atom systems with fixed Hamiltonians. Transition-metal systems introduce additional complexity through open-shell electronic structure, near-degenerate spin manifolds, and environment-dependent charge-transfer processes.

In this work, we apply SQD to the gas-phase and implicit-solvent potential energy surfaces of the synthetic test system \ce{[Co(H2O)5CO2]^{2+/3+}}. The complex is chosen as a controlled, chemically interpretable test system rather than a model of a specific catalyst. This system combines open-shell cobalt $d$-orbital physics, two accessible oxidation states, four spin multiplicities, and a clean metal-ligand dissociation coordinate at qubit counts tractable on current hardware. We examine the \ce{Co(III)} singlet and quintet states of \ce{[Co(H2O)5CO2]^{3+}} and the \ce{Co(II)} doublet and quartet states of \ce{[Co(H2O)5CO2]^{2+}} along the $\mathrm{Co-O_{(CO_2)}}$ coordinate. Active spaces are selected by atomic valence active-space (AVAS) projection~\cite{Sayfutyarova2017}. The charge-$+2$ doublet and quartet states, which exhibit chemically well-behaved dissociation, serve as controls and are evaluated at a $17$-orbital ($34$-qubit) active space. For the charge-$+3$ singlet and quintet states, where the chemistry of interest is concentrated, we run a systematic active-space convergence study through $(22e,17o)$, $(24e,19o)$, $(28e,23o)$, and $(30e,25o)$ - corresponding to $34$, $38$, $46$, and $50$ system qubits, respectively, under the Jordan-Wigner mapping, with additional auxiliary qubits allocated to mediate the density-density Jastrow couplings of the LUCJ ansatz. All hardware sampling is performed on IBM's Heron R3 processor \textsf{ibm\_pittsburgh}. To our knowledge, this is the first SQD hardware demonstration of an open-shell transition-metal complex with discrete metal-centered $d$-orbital chemistry. 

To probe environment-dependent electronic structure, the charged open-shell solute is placed in an IEF-PCM dielectric continuum, introducing a Hamiltonian that depends self-consistently on the correlated electronic density. Coupling SQD to IEF-PCM is structurally non-trivial because the inner SQD procedure inherently operates at fixed Hamiltonian, whereas the self-consistent reaction field condition makes the one-electron integrals a functional of the correlated density and therefore requires the entire self-consistent configuration recovery loop to be wrapped inside an outer self-consistent reactant field iteration~\cite{Kaliakin2025}. Extending the closed-shell SQD-IEF-PCM workflow of Kaliakin and co-workers to arbitrary $(N_\alpha, N_\beta)$ then follows from a restricted open-shell Hartree-Fock reference, a modified S-CORE step that preserves $\hat S_z \neq 0$, and an $\hat S^2$ constraint on the projected eigenproblem; together with the larger benchmark, it constitutes the first open-shell SQD calculation in implicit solvent.

Three principal findings emerge from the benchmark and dissociation-curve analyses presented here. First, across all four spin states and all four active-space sizes, SQD bitstrings sampled on \textsf{ibm\_pittsburgh} reproduce CCSD/UCCSD and heat-bath CI benchmarks at fixed active space and a shared Hamiltonian in both gas phase and implicit solvent, establishing methodological reliability for transition-metal $d$-orbital chemistry at the qubit counts targeted here. Second, the gas-phase \ce{[Co(H2O)5CO2]^{3+}} quintet exhibits an anomalous repulsive feature in its dissociation curve that is absent from the corresponding singlet and from both charge-$2$ states; Mulliken population analysis on one-particle density matrices indicates that this feature is consistent with an internal charge-transfer crossover from a localized $\mathrm{\ket{\{Co(H_2O)_5\}^{+3}~~CO_2}}$ description toward a charge-separated $\mathrm{\ket{\{Co(H_2O)_5\}^{+2}~~CO^+_2}}$ state that the open-shell quintet manifold makes spin-allowed. Third, IEF-PCM solvent stabilizes the charge-separated configuration and quenches the gas-phase repulsion, demonstrating that the open-shell SQD-IEF-PCM workflow correctly tracks environment-dependent stabilization of charged states. Together, these results establish SQD as a viable quantum-centric approach to transition-metal chemistry in regimes where spin state ordering, charge transfer, and environmental response are simultaneously active.

This article is organized as follows. Section~\ref{sec:methods} presents the computational methodology, emphasizing the elements specific to the present study: the open-shell mean-field treatment and its stability analysis, the AVAS active-space construction, the classical CCSD/UCCSD and HCI benchmarks, and the open-shell SQD and SQD–IEF-PCM workflows. Complete formulations of these methods, with all supporting equations, are collected in Supporting Information (SI). The results and their discussion are presented in Section~\ref{sec:results}.

\section{Methods}\label{sec:methods}
\subsection{System}

The left panel in Figure \ref{fig:main_figure} shows the Co-Complex as well as the low- and high-spin electronic configurations of \ce{Co(II)} and \ce{Co(III)} being in an octahedral field of point charges. The equilibrium geometry at a distance of \SI{1.97}{\angstrom} between the Co atom and the closest oxygen atom of \ce{CO2} was obtained for \ce{[Co(H2O)5CO2]^{3+}} in the Singlet spin state using the Geometric optimization solver within PySCF \cite{Sun2020PySCF} at HF level of theory and the def2-tzvp basis set. The corresponding XYZ coordinate file is included in the repository mentioned in the Data Availability section. All other geometry files are generated from this one by changing the distance between the Co atom and the \ce{CO2} molecule of the complex. The same geometry files are used for the examined oxidation and spin states (shown in Figure \ref{fig:main_figure}).\\

%%%%%%%%%%%%%%%%%%%%%%%%%%%%%%%%%%%%%%%%%%%%%%%%%%%%%%%%%%%%
\subsection{RHF \& ROHF}
\label{subsec:rhf_rohf}
Restricted Hartree-Fock (RHF) reference wave functions were used for the singlet state, whereas restricted open-shell Hartree-Fock (ROHF) references were employed for the doublet, quartet, and quintet states. Density fitting, also known as the resolution-of-the-identity approximation, was applied to the two-electron integrals using the auxiliary basis functions available in PySCF for the def2-TZVP basis set~\cite{Baerends1973ChemPhys,Whitten1973JCP,Dunlap1979JCP}. Initial RHF and ROHF calculations revealed multiple mean-field solutions along the dissociation coordinate. Internal stability analysis and, where required, second-order SCF optimization were therefore performed. The converged one-particle density matrix was propagated between adjacent geometries to obtain smooth and internally stable reference curves. Further details of the stability analysis and second-order SCF procedure as well as uncorrected mean-field curves are provided in SI~Sections~S1.1 and~S2. For the open-shell states, the ROHF references preserve the target spin sector required for the subsequent classical quantum chemistry methods as well as for SQD.

%%%%%%%%%%%%%%%%%%%%%%%%%%%%%%%%%%%%%%%%%%%%%%%%%%%%%%%%%%%%
\subsection{AVAS}
\label{subsec:avas}
Active spaces were constructed using the atomic valence active-space (AVAS) method~\cite{Sayfutyarova2017}. 
In AVAS, the occupied and virtual molecular-orbital spaces are projected separately onto a chemically selected atomic-valence subspace. The resulting projected overlap matrices are diagonalized within the occupied and virtual spaces, and the eigenvectors whose corresponding eigenvalues exceed specified thresholds are used to transform the original molecular orbitals. The resulting transformed occupied and virtual orbitals are retained in the active space. The target atomic orbitals were chosen to represent the Co-centered and CO$_2$-centered valence orbitals relevant to metal-ligand bonding. To control the numbers of occupied and virtual active orbitals independently, separate projection thresholds, $\tau_{\mathrm{occ}}$ and $\tau_{\mathrm{vir}}$, were applied to the occupied and virtual orbital spaces, respectively. This procedure enabled the systematic construction of the active spaces examined in this work while preserving their chemically relevant orbital character. The complete AVAS formulation, including the projection and orbital-transformation equations, and the thresholding procedure, is provided in SI~Section~S1.2.

%%%%%%%%%%%%%%%%%%%%%%%%%%%%%%%%%%%%%%%%%%%%%%%%%%%%%%%%%%%%
\subsection{CCSD}
\label{subsec:ccsd}

Coupled-cluster calculations with single and double excitations~\cite{Cizek1969,PurvisBartlett1982,ScuseriaSchaefer1989} were employed as classical benchmarks. Restricted CCSD was used for the singlet state, whereas unrestricted CCSD (UCCSD) was used for the doublet, quartet, and quintet states. The leading contribution from connected triple excitations was included through the perturbative triples correction, yielding CCSD(T) and UCCSD(T) for closed- and open-shell, respectively. For consistency with SQD and HCI, all single, double, and perturbative triple excitations were restricted to the occupied and virtual orbitals of the corresponding AVAS active space. Thus, the coupled-cluster, HCI, and SQD calculations were evaluated using matched active orbital spaces and the same underlying active-space Hamiltonians. The coupled-cluster ansatz, amplitude equations, and details of the active-space restriction are provided in  SI~Section~S1.3.

\subsection{HCI - gas and solvent phase} \label{subsec:hci}
Heat-Bath Configuration Interaction (HCI) was used as a selected configuration-interaction benchmark.
HCI constructs a compact variational determinant space by selecting external determinants according to the magnitude of their Hamiltonian coupling to the current wave function, followed by a second-order perturbative correction for determinants outside the variational space~\cite{Holmes2016}. In the gas phase, HCI was applied to the fixed active-space Hamiltonian obtained after integrating out the inactive doubly occupied orbitals. For implicit-solvent calculations, HCI was embedded in an IEF-PCM self-consistent reaction-field loop. At each iteration, the active-space one-electron integrals were updated using the reaction-field potential generated from the preceding correlated HCI one-particle reduced density matrix, whereas the two-electron integrals were kept fixed at their in vacuo values. The determinant-selection criterion, perturbative correction, active-space Hamiltonian, and correlated HCI-IEF-PCM iteration are described in detail in SI~Section~S1.4.

%%%%%%%%%%%%%%%%%%%%%%%%%%%%%%%%%%%%%%%%%%%%%%%%%%%%%%%%%%%%
\subsection{Sample-Based Quantum Diagonalization} 
\label{sec:sqd}

Sample-based quantum diagonalization (SQD)~\cite{RobledoMoreno2025} is a hybrid quantum-classical approach in which determinants sampled from an approximate quantum state define a subspace for classical diagonalization of the molecular Hamiltonian. In this work, the sampled state is prepared using the local unitary cluster Jastrow (LUCJ) ansatz~\cite{Motta2023},
\begin{equation}
  \lvert \Phi_{\mathrm{qc}} \rangle
  = e^{\hat{K}_{2}}\,e^{i\hat{J}_{1}}\,e^{\hat{K}_{1}}\,
    \lvert \mathbf{x}_{\mathrm{ref}} \rangle ,
  \label{eq:lucj}
\end{equation}
where $\hat{K}_{1}$ and $\hat{K}_{2}$ are one-body orbital rotations, $\hat{J}_{1}$ is a density--density Jastrow operator, and $\lvert \mathbf{x}_{\mathrm{ref}}\rangle$ is the Hartree--Fock reference determinant under the Jordan--Wigner mapping. The LUCJ parameters are obtained without variational optimization from the $t_1$ and $t_2$ amplitudes of classical CCSD/UCCSD calculations in the corresponding active space through double factorization~\cite{RobledoMoreno2025,Kaliakin2025}.

The measured bitstrings are processed using self-consistent configuration recovery (S-CORE)~\cite{RobledoMoreno2025}. S-CORE restores the target particle-number and $S_z$ sectors, constructs batches of recovered determinants, and iteratively updates the orbital occupations from the one-particle reduced density matrices obtained after projected Hamiltonian diagonalization. The resulting determinant subspaces are diagonalized classically to obtain the SQD energies and correlated densities.

For the open-shell states considered here, we follow the formulation of Ref.~\cite{Liepuoniute2025}. When $N_\alpha \neq N_\beta$, configuration recovery is performed separately in the $\alpha$- and $\beta$-spin registers so that the recovered determinants retain the target $(N_\alpha,N_\beta)$ electron numbers and therefore remain in the desired $S_z$ sector. The spin-inversion symmetrization employed for closed-shell SQD is consequently disabled (\texttt{symmetrize\_spin=False}). The remaining sampling, configuration-recovery, batching, and projected-diagonalization steps follow the standard SQD workflow~\cite{RobledoMoreno2025}. Further details on the SQD workflow are provided in SI~Section~S1.5.

%%%%%%%%%%%%%%%%%%%%%%%%%%%%%%%%%%%%%%%%%%%%%%%%%%%%%%%%%%%%
\subsection{IEF-PCM for Open-Shell Solutes}
\label{sec:iefpcm}

Solvent effects are described using the integral-equation-formalism polarizable continuum model (IEF-PCM)~\cite{Tomasi2005,Mennucci1997,Cances1997}, in which the solute is embedded in a polarizable dielectric continuum. The solvent response enters the electronic Hamiltonian through a density-dependent one-electron reaction-field potential. Because this potential depends on the solute electronic density, the correlated electronic structure and solvent polarization must be determined self-consistently through a self-consistent reaction-field (SCRF) procedure.

For an open-shell solute, the IEF-PCM response is determined by the total electronic density,
$\rho=\rho_\alpha+\rho_\beta$, rather than by the spin density. We use an IEF-PCM ROHF solution for the non-singlet states, providing a spin-pure reference determinant in the target $S$ and $S_z$ sector. In the correlated calculations, the solvent contribution is incorporated through the density-dependent one-electron operator, whereas the two-electron integrals are retained at their in vacuo values~\cite{Caricato2012,Cammi2009,Caricato2018,Castaldo2022} (the frozen-reaction-field approximation). Additional details of the IEF-PCM formulation are provided in SI~Section~S1.6.

%%%%%%%%%%%%%%%%%%%%%%%%%%%%%%%%%%%%%%%%%%%%%%%%%%%%%%%%%%%%
\subsection{Open-Shell SQD with IEF-PCM}
\label{sec:sqd-iefpcm}

We extend the closed-shell SQD--IEF-PCM workflow of Kaliakin \textit{et al.}~\cite{Kaliakin2025} to open-shell active spaces through three modifications: an ROHF reference determinant, an $S_z$-preserving S-CORE procedure, and an $\hat S^2$ constraint on the projected eigenproblem. The SQD procedure of Section~\ref{sec:sqd} is then embedded within an outer self-consistent reaction-field (SCRF) loop that updates the solvent reaction field from the correlated SQD density.

At SCRF iteration $k$, the active-space one-electron integrals are updated according to
\begin{equation}
  h_{pq}^{(k)}
  =
  h_{pq}^{(0)}
  +
  V_{pq}^{\mathrm{solv}}\!\left[\gamma^{(k-1)}\right],
  \label{eq:hcore-update-main}
\end{equation}
where $h_{pq}^{(0)}$ are the in vacuo one-electron integrals and
$V_{pq}^{\mathrm{solv}}[\gamma^{(k-1)}]$ is the IEF-PCM reaction-field contribution generated from the correlated total one-particle density matrix of the preceding SCRF iteration. The two-electron integrals remain fixed at their in vacuo values. An inner SQD/S-CORE calculation with the updated Hamiltonian provides the correlated energy and density used in the subsequent SCRF cycle.

For each geometry and spin state, the solvated circuit is sampled once on the QPU. The SCRF iterations then reuse these bitstrings, so solvent self-consistency is achieved entirely in classical post-processing without additional hardware sampling. Further details of the open-shell symmetry treatment, solvent-energy functional, and SCRF implementation are provided in SI~Section~S1.7.

%%%%%%%%%%%%%%%%%%%%%%%%%%%%%%%%%%%%%%%%%%%%%%%%%%%%%%%%%%%%
\subsection{Computational Details}
\label{sec:comp-details}
\paragraph{Mean-field reference and active space.}
% Mean-field reference wavefunctions were obtained using RHF for singlet and ROHF calculations anywhere else, in \texttt{PySCF} with the def2-TZVP basis set. 
For implicit solvent, ROHF was combined with an IEF-PCM model (dielectric constant $\varepsilon = 78.3553$) using density fitting and a second-order SCF solver, with convergence threshold $10^{-9}$ and a maximum of 50 SCF cycles. Stability analysis and re-optimization of the density matrix were applied to ensure internally stable solutions. For gas-phase calculations, the same ROHF protocol was used without the solvent model.

Active spaces were constructed using the AVAS procedure targeting selected atomic orbital subspaces corresponding to chemically relevant valence orbitals. The resulting AVAS orbitals were reordered and subsequently processed using a blockwise rank-fixing procedure that enforces a fixed active-space size while preserving the occupied/virtual character. For both phases, the representative sizes ranged from $(N_{\mathrm{orb}}, N_\alpha, N_\beta) = (17$-$25,\,11$-$17,\,9$-$15)$. The corresponding number of active electrons was determined consistently from the ROHF reference occupation and spin ($N_\alpha$, $N_\beta$), ensuring proper open-shell configurations.

\paragraph{Classical correlated benchmarks (gas phase).}
% Classical correlated reference calculations were performed using coupled cluster and selected configuration interaction methods within the same active spaces. 
CCSD(T) energies were obtained using \texttt{PySCF} with default convergence settings and no additional approximations. HCI calculations were carried out in deterministic mode, using selection and perturbative thresholds of $\epsilon_1 = 10^{-3}$ and $\epsilon_2 = 10^{-5}$, respectively.

\paragraph{Classical correlated benchmarks (implicit solvent).}
Classical correlated reference calculations in implicit solvent were performed using the same active-space Hamiltonians within a self-consistent reaction-field (SCRF) framework implemented in \texttt{PySCF}. CCSD(T) energies were obtained using default settings on top of ROHF references including the solvent potential. Heat-Bath Configuration Interaction (HCI) calculations were carried out in deterministic mode with thresholds $\epsilon_1 = 10^{-3}$ and $\epsilon_2 = 10^{-5}$, embedded within an SCRF loop in which the one-electron integrals were iteratively updated from the density-dependent solvent contribution, while two-electron integrals were kept fixed. The SCRF cycle was repeated 4 times. %, achieving convergence of the total energy of \~$10^{-5}$~Ha. 

\paragraph{LUCJ circuit construction.}
For each geometry, UCCSD amplitudes $(t_1,t_2)$ obtained from classical calculations were loaded and mapped onto the LUCJ ansatz using the \texttt{ffsim.UCJOpSpinUnbalanced} operator. The interaction structure was restricted to local fermionic couplings: nearest-neighbour interactions along same-spin orbital chains ($( \alpha\alpha )$ and $( \beta\beta )$ blocks) and sparse opposite-spin density-density couplings ($( \alpha\beta )$) introduced periodically (every fourth spatial orbital), reflecting the hardware-aware reduction of the full Jastrow connectivity.  

The quantum register consisted of $2N_{\mathrm{orb}}$ qubits corresponding to spin orbitals under the Jordan-Wigner mapping. Circuits were initialized in the Hartree-Fock reference determinant using \texttt{ffsim}\texttt{.qiskit}\texttt{.Prepare- }\texttt{HartreeFockJW}, followed by application of the LUCJ unitary via \texttt{ffsim.qiskit.UCJOpSpinUnbalancedJW}. All qubits were measured in the computational basis to produce bitstrings for SQD sampling.

To enable efficient execution on IBM heavy-hex architectures, a custom “zig-zag” layout was constructed using rustworkx, embedding the $\alpha$ and $\beta$ spin-orbital chains as two parallel linear subgraphs connected by auxiliary mediator qubits inserted periodically (every fourth orbital) to realize the reduced $(\alpha\beta)$ couplings while preserving subgraph isomorphism with the backend connectivity. 
Among all valid embeddings, the initial layout was selected using a heuristic cost function minimizing two-qubit gate errors and readout errors. Circuits were then transpiled using Qiskit preset pass managers (optimization level 3), with additional fermion-aware pre-initialization passes (\texttt{ffsim.qiskit.PRE\_INIT}) applied prior to routing. 

\paragraph{Quantum hardware execution and error handling.}
All circuits were executed on IBM Heron R3 processor \texttt{ibm\_pittsburgh} using Q-CTRL’s Fire Opal software \cite{qctrl_fireopal}, including circuits reduction step of the QPU-transpiled circuits. Q-CTRL's Fire Opal provides automated error suppression through hardware-aware transpilation, calibration-aware gate scheduling, and measurement optimization, improving circuit fidelity without introducing sampling overheads; in particular, each shot directly yields a physical bitstring used in the SQD workflow. Sampling budgets were scaled with system size: $2\times10^5$ shots for 34 qubits, $3\times10^5$ shots for 38 qubits, and $5\times10^5$ shots for 46- and 50-qubit circuits. These shot counts were chosen to balance statistical convergence of the sampled determinant distribution with hardware runtime constraints. Further details on circuit optimization, execution configuration, and raw QPU performance analysis are provided in SI~Sections~S3 and~S4.

\paragraph{SQD post-processing (gas phase).}
The active-space Hamiltonian that was then subspace projected in the post-processing routine, was defined by one- and two-electron integrals $(h_{pq}, g_{pqrs})$ obtained from ROHF-based AVAS calculations. S-CORE procedure was implemented using \texttt{qiskit-addon-sqd} workflow. Per each data point it was run on single HPC node with 3 batches with $2\times10^3$ samples per batch, with maximum number of 50 iterations and convergence thresholds of $10^{-5}$~Ha in energy and $10^{-9}$ in occupancies. 

\paragraph{SQD post-processing (implicit solvent).}
For solvent calculations, the SQD procedure was embedded in an outer self-consistent reaction-field (SCRF) loop, coupling the SQD-derived electronic density to a continuum solvation model implemented in \texttt{PySCF}. The quantum-sampled bitstrings were reused across all SCRF iterations, with no additional hardware sampling required. Each SCRF iteration consisted of an inner SQD loop using the S-CORE procedure with adaptive iteration counts (ranging from $2$ to $15$ inner S-CORE iterations, per SCRF cycle), controlled by energy and occupancy thresholds of $10^{-5}$~Ha and $10^{-4}$, respectively. The determinant space was partitioned into $N_{\mathrm{batch}}=3$ batches with $\sim 1.5\times10^3$ samples per batch, and the projected Hamiltonian was diagonalized as in the gas-phase case, including a soft $S^2$ constraint where required. Following each inner loop, the one-electron integrals were updated using the density-dependent solvent potential, while two-electron integrals were kept fixed (frozen-reaction-field approximation). Convergence of the SCRF loop was typically achieved within $5$-$6$ iterations, with progressively tightened inner-loop parameters. 

\paragraph{Software}
Classical quantum chemistry simulations were done in PySCF\cite{Sun2018} 2.9.0 with SHCISCF interface\cite{pyscf_shciscf_2021} 0.1 for Heat-Bath CI calculations. 
Quantum computing part of workflow was implemented using Qiskit\cite{Qiskit} 2.2.1 extended with Qiskit Nature\cite{qiskit_nature_2023} 0.7.2, ffsim\cite{sung2026ffsim} 0.0.59 and Qiskit-SQD-Addon\cite{qiskit-addon-sqd} 0.12.0.

\paragraph{Hardware}
SQD postprocessing and HCI solvers were run at the PSNC's HPC cluster Eagle. 
Each single data point post-processing/ground-state estimation with HCI was run on a single node with varied parameters with up to ~200GB RAM and 96-threads CPU, no GPU parallelization.

%%%%%%%%%%%%%%%%%%%%%%%%%%%%%%%%%%%%%%%%%%%%%%%%%%%%%%%%%%%%
\section{Results}
\label{sec:results}

\begin{table}[H]
    \caption{AVAS target AOs and active space sizes for $[\mathrm{Co(H_2O)_5CO_2}]^{3+}$. The listed AO labels belong to \ce{Co} and \ce{CO2} of the complex exclusively.}
    \label{tab:active-spaces}
    \centering
    \begin{tabular}{|l|c|c|c|c|c|}
        \hline
        Target AOs    & $n_{\textrm{mo}}$ & $n_{\textrm{qubits}}$ & $n_{\textrm{elec}}$ & Singlet $(n_{\alpha},n_{\beta})$ & Quintet $(n_{\alpha},n_{\beta})$ \\ \hline\hline
        \{Co 3d, 4s, 4p;   & 17 & 34 & 22 & (11,11)& (13,9)  \\
          ~~O 2s, 2p;        & 19 & 38 & 24 & (12,12)& (14,10) \\
          ~~C 2s, 2p;        & 23 & 46 & 28 & (14,14)& (16,12)  \\
          ~~O 2s, 2p\}       & 25 & 50 & 30 & (15,15)& (17,13)\\ \hline
    \end{tabular}
\end{table}

\subsection{Gas phase energetics}
\label{sec:gas-results}

\subsubsection{Equilibrium spin state energetics for $\mathrm{\ce{[Co(H2O)5CO2]^{3+}}}$}
We want to first evaluate spin state energetics using SQD at the equilibrium geometry of $\mathrm{\ce{[Co(H2O)5CO2]^{3+}}}$ ($r_{\mathrm{Co\!-\!O(CO_2)}}=1.97$~\AA) across the two spin states and the four active-space sizes. For the charge-$+3$ singlet ($S=0$) and quintet ($S=2$), we perform our calculations on active spaces with 17, 19, 23 and 25 molecular orbitals corresponding to 34, 38, 46 and 50 spin-orbitals respectively. Under Jordan-Wigner mapping, these active spaces correspond to 34, 38, 46, and 50 system qubits. The total electron count is matched between the singlet and quintet at each active space size so that the spin state gap is computed at strictly consistent active spaces. Table~\ref{tab:active-spaces} summarizes the active space configurations, number of qubits, and electron counts in each spin configuration for the $\mathrm{\ce{[Co(H2O)5CO2]^{3+}}}$.

\begin{figure}[H]
    \centering
    \includegraphics[width=0.47\columnwidth]{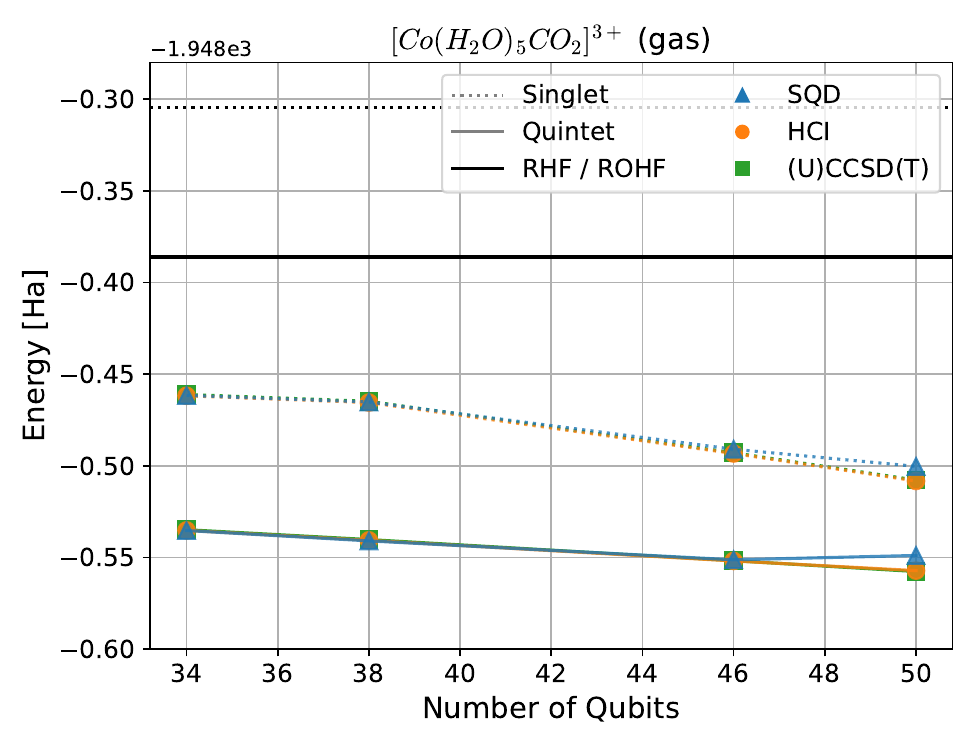}
    \caption{ Gas phase equilibrium energies: RHF/ROHF / CCSD(T)/UCCSD(T) / HCI / SQD energies on y-axis at equilibrium geometry vs qubit count corresponding to the active spaces considered on x-axis for +3 charge singlet and quintet.}
    \label{fig:eq-vs-qubits_gas_charge_3}
\end{figure}

Across the four active-space sizes (17, 19, 23, and 25 orbitals), for both spin states (singlet and quintet) the SQD energies sampled on \texttt{ibm\_pittsburgh} after self-consistent configuration recovery convergence track the CCSD(T)/UCCSD(T) reference within 0.51, 0.72, 1.93, and 8.72~m$E_h$, respectively and for the comparison the variational HCI benchmark track the CCSD(T)/UCCSD(T) within 0.69, 0.67, 0.40, 0.50~m$E_h$ at the equilibrium geometry, while the bare ROHF energy lies several tens of m$E_h$ above the all correlated methods. The hierarchy $E_{\mathrm{ROHF}} \gg E_{\mathrm{CCSD/UCCSD}} \approx E_{\mathrm{HCI}} \approx E_{\mathrm{SQD}}$ holds uniformly for the four states and is illustrated as a function of qubit count in Figure~\ref{fig:eq-vs-qubits_gas_charge_3}. The fact that SQD - whose subspace is constructed from bitstrings drawn from a noisy LUCJ circuit, parameterised without further variational optimisation from the classical CCSD $t_1$, $t_2$ amplitudes via double factorization - agrees with HCI to within 0.19, 0.30, 2.33, 8.27~m$E_h$ (respectively for 17, 19, 23, and 25 orbitals) is non-trivial for an open-shell $3d$ problem at $(N_\alpha\!\neq\!N_\beta)$. This includes the largest $(30e,25o)$ active space, in which the active-space FCI dimension already exceeds $\sim\!10^{12}$.

The chemically meaningful observables are the spin state gaps $\Delta E_{\mathrm{singlet-quintet}}$ for charge $+3$, reported in Table~\ref{tab:eq_gaps_gas_charge_3}. Over all of the active space sizes SQD gap reproduces the HCI gap, while differing by no more than 1.07~kcal\,mol$^{-1}$, and within 0.90~kcal\,mol$^{-1}$ of CCSD(T). Further we observe that HF, CCSD(T)/UCCSD(T), HCI and SQD show that the Quintet high-spin state is energetically lower than the Singlet state, as is expected for a $\mathrm{3d^6}$ electronic configuration with weak ligand caused energetic splitting between the $e_g$ and $t_{2g}$ orbitals.

\begin{table}[H]
    \caption{Singlet-quintet energy gaps $\Delta E = E(\mathrm{singlet}) - E(\mathrm{quintet})$ at equilibrium as a function of active-space size 
    $(n_{\textrm{elec}}, n_{\textrm{mo}})$ in the gas phase for charge=3. Results from ROHF, CCSD(T) for singlet and UCCSD(T) for non-singlet, CCSD(T) for singlet, HCI, and SQD are compared, highlighting the dependence of correlation treatment on the chosen active space.}
    \label{tab:eq_gaps_gas_charge_3}
    \centering
    \begin{tabular}{|c|l|l|l|l|}
        \hline
        $\Delta E_{\textrm{ROHF}}\textrm{/mHa}$ 
        & $(n_{\textrm{elec}}, n_{\textrm{mo}})$ 
        & $\Delta E_{\textrm{CCSD(T)}}\textrm{/mHa}$ 
        & $\Delta E_{\textrm{HCI}}\textrm{/mHa}$ 
        & $\Delta E_{\textrm{SQD}}\textrm{/mHa}$ \\ \hline
        \multirow{4}{*}{81.7}
        & (22, 17) & 76.2 & 73.4 & 73.7 \\ \cline{2-5}
        & (24, 19) & 78.4 & 75.2 & 75.8 \\ \cline{2-5}
        & (28, 23) & 62.9 & 58.6 & 60.8 \\ \cline{2-5}
        & (30, 25) & 54.3 & 48.9 & 40.4 \\ \hline
    \end{tabular}
\end{table}
\FloatBarrier

\begin{figure}[H]
    \centering
    \includegraphics[width=1\linewidth]{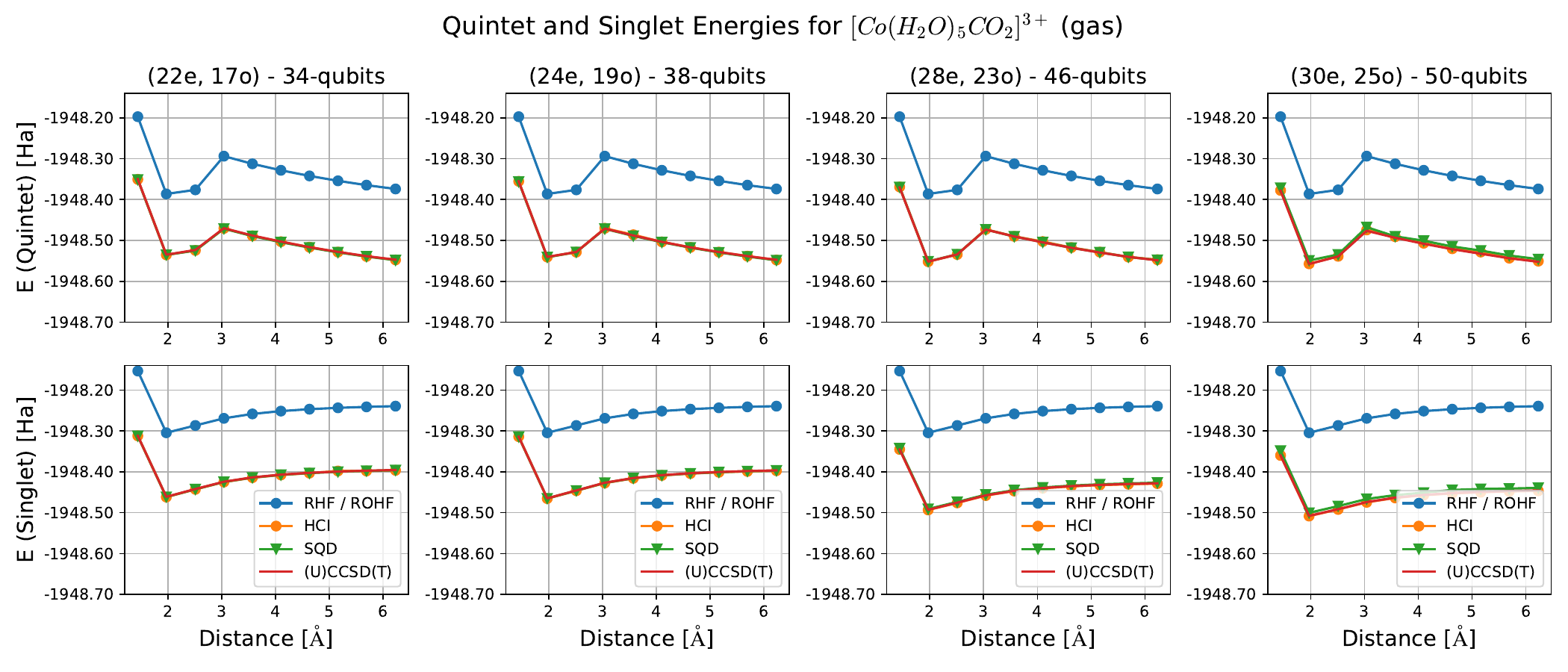}
    \caption{Potential energy curves for the quintet (top row) and singlet (bottom row) states of [Co(H$_2$O)$_5$CO$_2$]$^{3+}$ in the gas phase, computed for increasing active-space sizes. Results from RHF/ROHF, HCI, (U)CCSD(T), and SQD are compared as a function of the $\mathrm{Co-O_{(CO_2)}}$ distance.}
    \label{fig:gas_charge_3_results}
\end{figure}

\FloatBarrier

\subsubsection{Charge-$+3$ singlet dissociation curve.}
We then trace the potential energy curve along the $\mathrm{Co-O_{(CO_2)}}$ coordinate for the closed-shell $\mathrm{Co(III)}$ singlet of $\mathrm{\ce{[Co(H2O)5CO2]^{3+}}}$ at the four active-space sizes (Figure~\ref{fig:gas_charge_3_results}). The curve is 
%monotonic - mathematically it is not monotonic
well-behaved: an attractive minimum near 1.97~\AA, a smooth rise as $\mathrm{r_{Co-O_{(CO_2)}}}$ is elongated, and a flat asymptote in the dissociation limit consistent with separated $\{\mathrm{Co(H_2O)_5}\}^{3+}+\mathrm{CO_2}$ fragments.

For the singlet case, across the entire coordinate, through active spaces (17, 19, 23, and 25 orbitals), the SQD curve overlays UCCSD(T)/CCSD(T) differing by no more than 0.57, 0.51, 2.84, and 11.55~m$E_h$, respectively, while SQD curve again overlays HCI differing by no more than 1.94, 0.37, 2.62, and 11.30~m$E_h$ and exhibits no discontinuities, kinks, or qualitative artefacts. Energy convergence with respect to qubit count is smooth and uniform: enlarging the active space from 34 to 50 qubits lowers the curve nearly rigidly, with the spacing between successive sizes decreasing systematically. We note here that the bare RHF curve is sensitive to multiple SCF solutions in the intermediate $\mathrm{r_{Co-O_{(CO_2)}}}$ region; smooth RHF curves were obtained only after second-order Newton SCF combined with one-particle density matrix propagation between adjacent geometries, as detailed in Methods (Sec. \ref{sec:methods}).

\begin{figure}
    \centering
    \includegraphics[width=0.5\columnwidth]{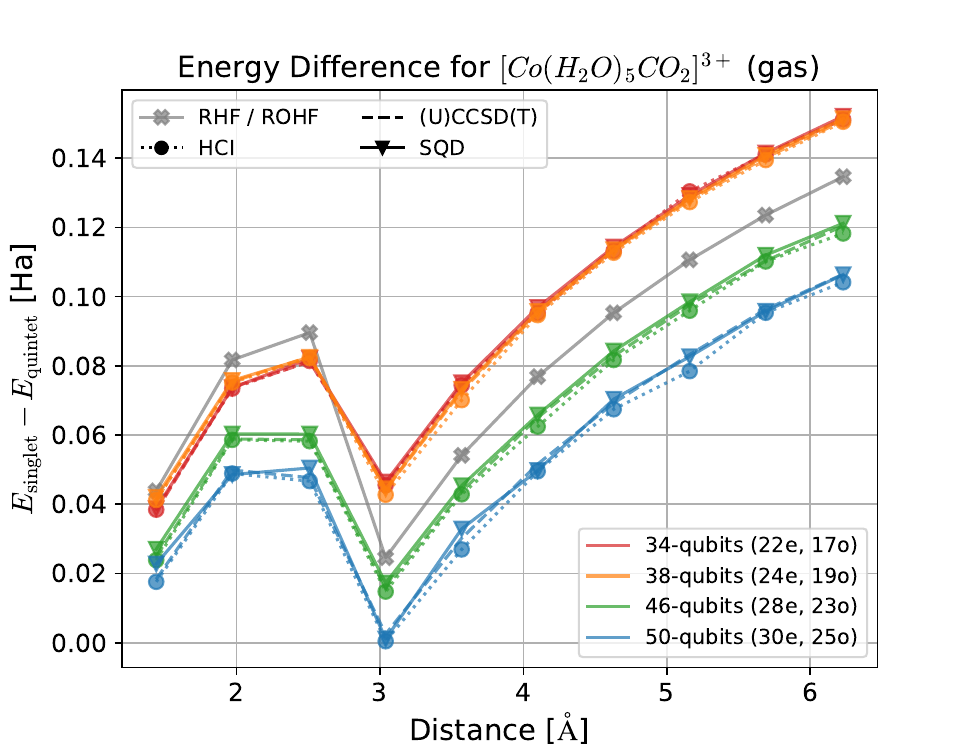}
    \caption{Singlet-quintet energy differences $\Delta E = E(\mathrm{singlet}) - E(\mathrm{quintet})$ for [Co(H$_2$O)$_5$CO$_2$]$^{3+}$ in the gas phase as a function of $\mathrm{Co-O_{(CO_2)}}$ distance. Results are shown for increasing active-space sizes and compared across ROHF, HCI, CCSD(T)/UCCSD(T), and SQD methods.}
    \label{fig:gas_charge_3_diff_results}
\end{figure}

\subsubsection{Charge-$+3$ quintet dissociation curve.}\label{sec:results-charge3-quintet}
\begin{figure}[h]
    \centering
    \includegraphics[width=0.5\columnwidth]{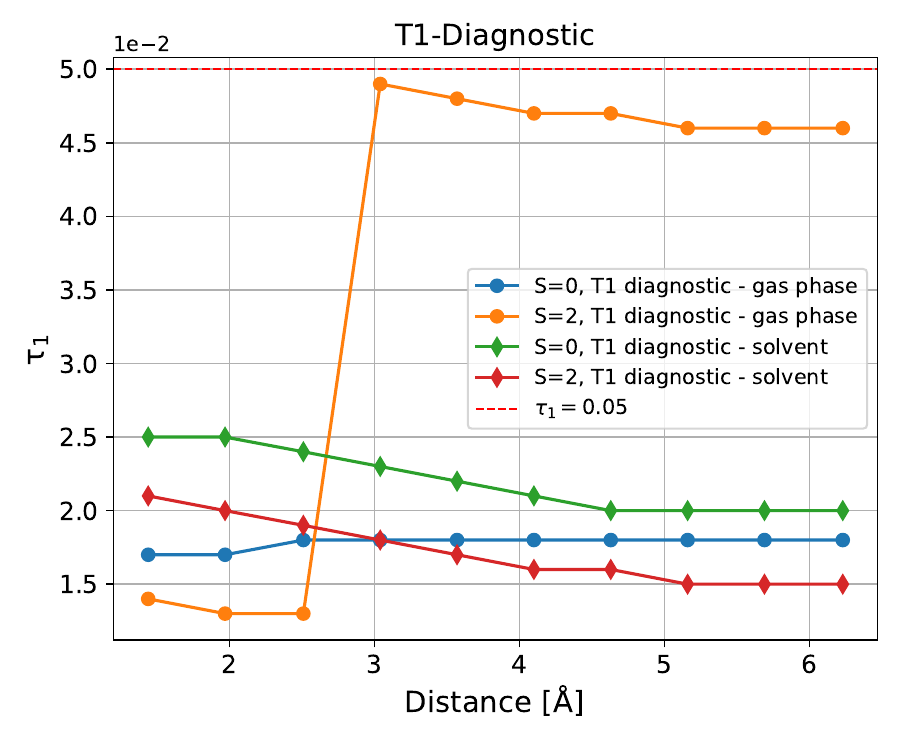}
    \caption{$\mathrm{\tau_1}$-diagnostics from CCSD/UCCSD calculations along the  $\mathrm{Co-O_{(CO_2)}}$ distance for the singlet (S=0) and quintet (S=2) states in gas phase and solvent for the (30e,25o) active space.}
    \label{fig:t1_diagnostic_singlet_quintet_gas_solvent}
\end{figure}

\begin{figure}[h]
    \centering
    \includegraphics[width=0.5\columnwidth]{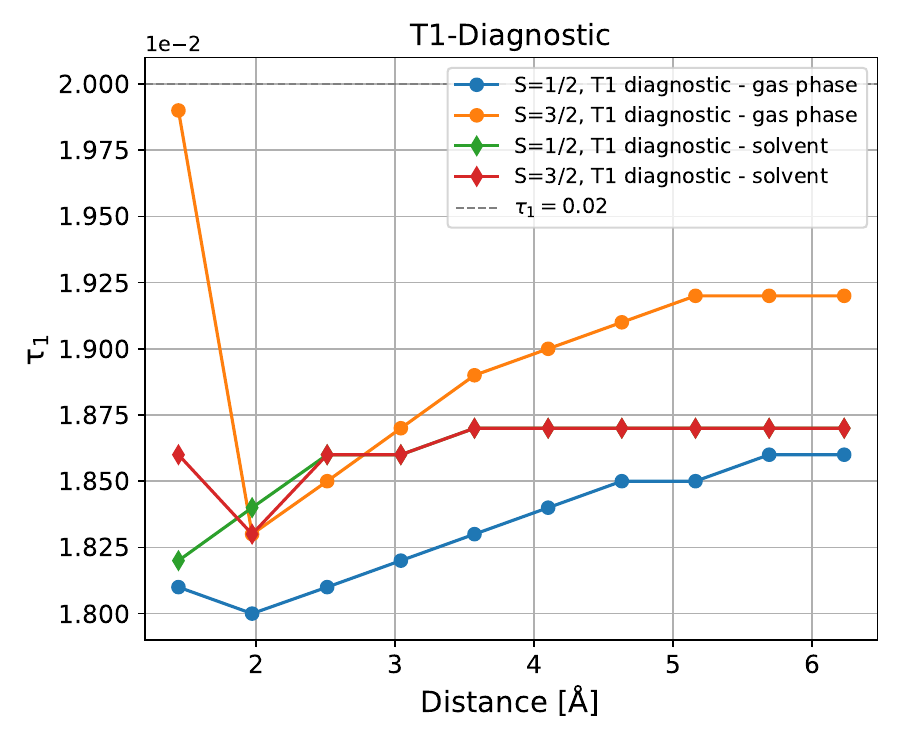}
    \caption{$\mathrm{\tau_1}$-diagnostics from UCCSD calculations along the  $\mathrm{Co-O_{(CO_2)}}$ distance for the doublet (S=1/2) and quartet (S=3/2) states in gas phase and solvent for the (23e,17o) active space.}
    \label{fig:t1_diagnostic_doublet_quartet_gas_solvent}
\end{figure}

The high-spin $\mathrm{Co(III)}$ quintet of $[\mathrm{Co(H_2O)_5CO_2}]^{3+}$ exhibits qualitatively different physics. The UCCSD(T)/ CCSD(T), HCI, and SQD curves all display a non-monotonic feature near $r_{\mathrm{Co\!-\!O(CO_2)}}\!\approx\!3$~\AA: the energy first decreases until the equilibrium minimum, then rises through a local maximum of magnitude 70-90~m$E_h$ before relaxing again (Figure~\ref{fig:gas_charge_3_results}).
The feature is present at all four active-space sizes (34, 38, 46, 50 qubits) and persists in HCI, ruling out an artefact of either subspace size or the SQD sampling. The ROHF curve is the most strongly distorted in this region. Stability analysis on the ROHF reference is performed on all points and the second-order Newton solver is required throughout this region to converge to true stationary points. For closed-shell molecules the so-called t1-diagnostics, the Euclidean norm of the singles amplitudes vector divided by the square root of the number of the correlated electrons, has been proposed by Lee and co-workers \cite{Lee1989TCA_FOOF,Lee1989IJQC_Diagnostic}. There a t1-diagnostics value larger than 0.02 indicates the presence of a multi-reference character. In reference \cite{David2020PhD} the t1-diagnostics for high-spin open-shell molecules within the unrestricted CC2 model has been defined as $\tau_1 = \sqrt{\frac{t^{\alpha}_1t^{\alpha}_1+ t^{\beta}_1t^{\beta}_1}{n_{\alpha}+n_{\beta}}}$. It is concluded that for alkane radicals values larger than 0.02 and for other doublet molecules values around 0.05 indicate a degenerate ground state that might be better represented by multi-reference methods. In this work the singles amplitudes and number of correlated electrons used to calculate $\tau_1$ are restricted to the active space size. For all the considered spin states the $\tau_1$ value is clearly under 0.03, except for the quintet case in gas phase where we see a sudden increase to $\approx$ 0.05 starting at the distance of 3.04~\AA~between $\ce{Co}$ and the closest oxygen of $\ce{CO2}$, see Figures \ref{fig:t1_diagnostic_singlet_quintet_gas_solvent} and \ref{fig:t1_diagnostic_doublet_quartet_gas_solvent}. As described below, this behavior is due to a charge separation in the system and is not present for the quintet state in the solvent, where the $\tau_1$ value again remains clearly below 0.03 along the potential energy curve.

\begin{figure}[h]
    \centering
    \includegraphics[width=0.5\columnwidth]{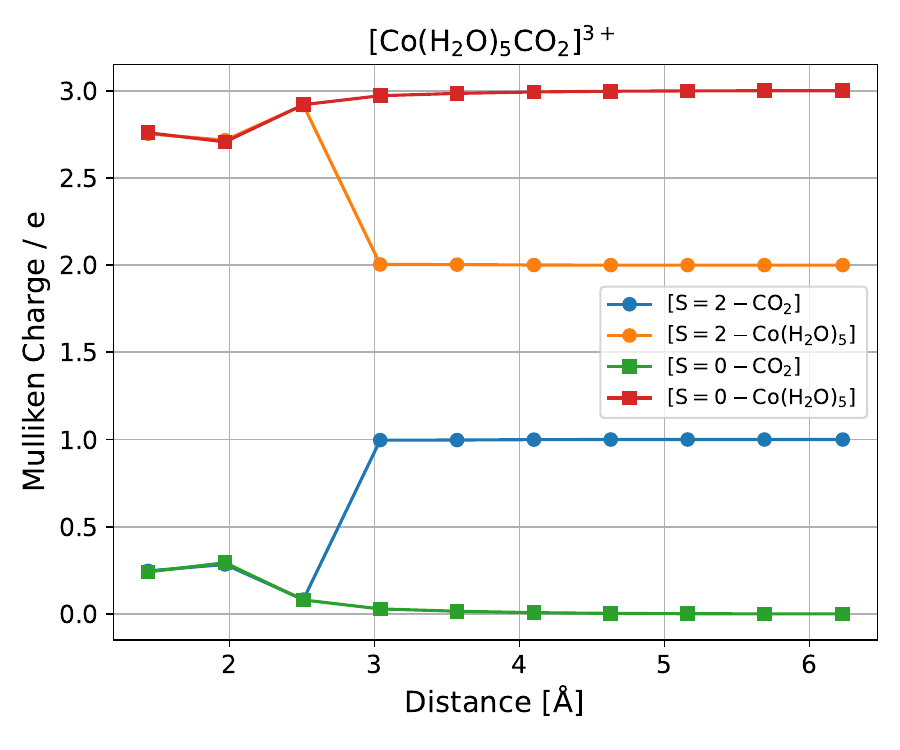}
    \caption{Mulliken population analysis from RHF/ROHF 1-RDM along the  $\mathrm{Co-O_{(CO_2)}}$ distance for the Singlet (S=0) and quintet (S=2) states in gas phase.}
    \label{fig:gas_mulliken_singelt_quintet}
\end{figure}

To identify the physics, we perform a Mulliken population analysis on the one-particle reduced density matrix $\rho_{pq} = \langle\psi_{ROHF}|\,\hat a_{p}^\dagger\hat a_{q}|\psi_{ROHF}\rangle$ extracted from the stable ROHF wave function, partitioned between the $\{\mathrm{Co(H_2O)_5}\}$ moiety and the CO$_2$ ligand, see figure \ref{fig:gas_mulliken_singelt_quintet}.

\begin{figure}[h]
    \centering
    \includegraphics[width=0.5\columnwidth]{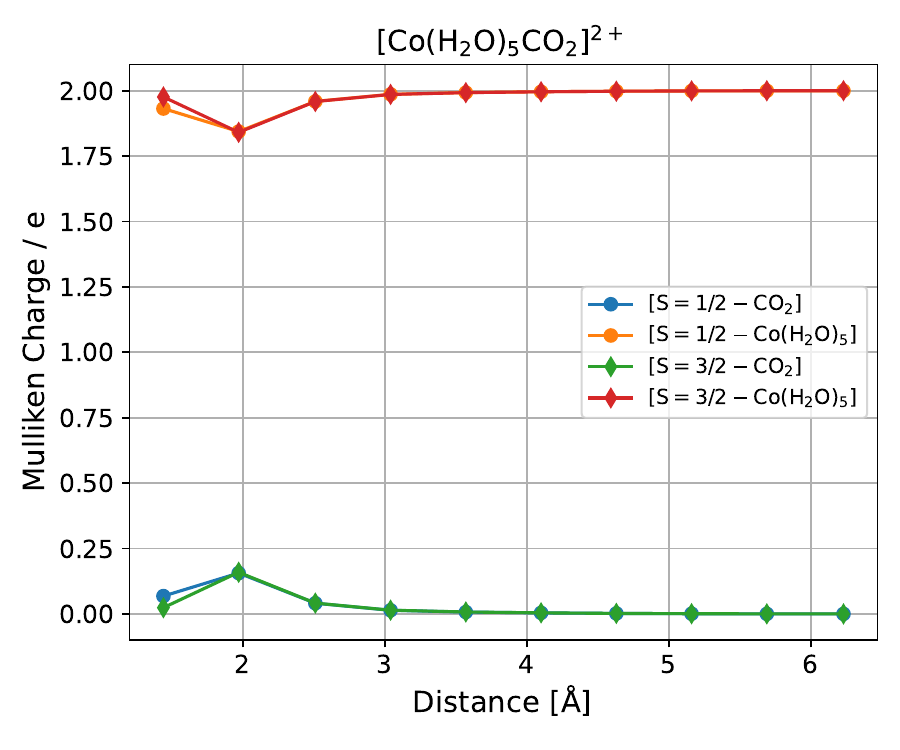}
    \caption{Mulliken population analysis from ROHF 1-RDM along the  $\mathrm{Co-O_{(CO_2)}}$ distance for the Doublet (S=1/2) and Quartet (S=3/2) states in gas phase.}
    \label{fig:gas_mulliken_doublet_quartet}
\end{figure}

At equilibrium the partial charges are close to $q[\mathrm{Co(H_2O)_5}]\!\approx\!+3$ and $q[\mathrm{CO_2}]\!\approx\!0$, i.e.\ a localised $\mathrm{\{Co(H_2O)_5\}^{+3}}\!+\!\mathrm{\{CO_2\}}$ description. On elongation, the Mulliken charges undergo a sharp redistribution near $r\!\approx\!3$~\AA: $q[\mathrm{Co(H_2O)_5}]$ drops by approximately one electron unit toward $\sim\!+2$ while $q[\mathrm{CO_2}]$ rises toward $\sim\!+1$, after which both quantities plateau. The spatial location of the Mulliken crossover coincides with the local maximum on the energy curve, identifying the feature as an avoided crossing between two diabats:
\begin{equation}
\big|\,\mathrm{\{Co(H_2O)_5\}^{+3}}\!~~~\!\mathrm{CO_2}\,\big\rangle
\;\longleftrightarrow\;
\big|\,\mathrm{\{Co(H_2O)_5\}^{+2}}\!~~~\!\mathrm{CO_2^+}\,\big\rangle .
\label{eq:CT-diabats}
\end{equation}
An internal electron-transfer from the $\{CO_2\}$ fragment to the $\mathrm{\{Co(H_2O)_5\}}$ fragment
occurs while the overall spin state with $S=2$ can be conserved by a high-spin coupling of a quartet open-shell on $\mathrm{\{Co(H_2O)_5\}}$ and a doublet open-shell on $\{CO_2\}$. This is not possible for the charge-+3 Singlet state with $\mathrm{Co(III)}\,d^6$ electronic configuration. 
Beyond the crossing, the residual rise in energy reflects the Coulombic repulsion between the $+2$ and $+1$ fragments along the same dissociation coordinate, which relaxes only slowly with $1/r$ in the gas phase.

The agreement of SQD with HCI through the crossing region is the more demanding test of the method.
SQD reproduces the position of the crossover, the height of the local maximum, and the post-crossing slope. Over the whole dissociation curve, SQD energies differ from HCI energies by no more than 1.39, 2.85, 1.57, 8.27~m$E_h$, respectively for 17, 19, 23, and 25 orbitals for four active-space sizes, demonstrating that the LUCJ ansatz initialised from gas-phase CCSD amplitudes, combined with S-CORE on the noisy bitstrings, produces a subspace that correctly spans both diabats in Eq.~\eqref{eq:CT-diabats} and their coupling. It is worth noting, that the larger deviations observed for the largest active space should be attributed to the limited number of S-CORE iterations performed. This limitation stems from the fixed 48-hour budget allocated to post-processing, which was maintained consistently across all calculations to ensure a compact and comparable post-processing procedure. Detailed numbers of post-processing iterations can be seen SI Section~S4. 

\FloatBarrier

\begin{figure}[H]
    \centering
    \includegraphics[width=0.5\linewidth]{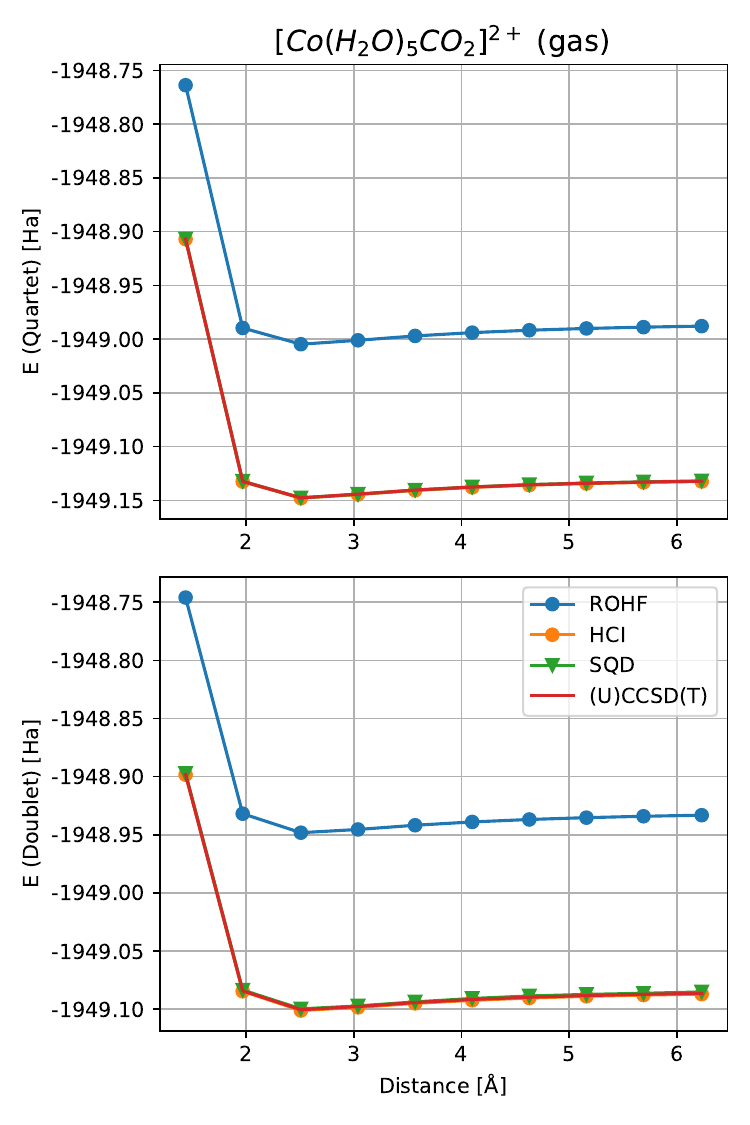}
    \caption{Figure shows potential energy curves for doublet and quartet for charge=2 for 34 qubits in the gas phase. Active space includes $(17o,12n_\alpha,11n_\beta)$ for doublet and  $(17o,13n_\alpha,10n_\beta)$ for quartet.}
    \label{fig:gas_charge_2_results}
\end{figure}

\subsubsection{Charge-$+2$ doublet and quartet.}
\label{sec:results-charge2}
As an internal control we run the doublet ($S=1/2$) and quartet ($S=3/2$) states of $[\mathrm{Co(H_2O)_5CO_2}]^{2+}$ at the 34-qubit $(M_o=17)$ active space along the same dissociation coordinate (Figure~\ref{fig:gas_charge_2_results}). Both curves are kink-free, and asymptote smoothly to separated $\{\mathrm{Co(H_2O)_5}\}^{2+}+\mathrm{CO_2}$ fragments. The Mulliken charges remain close to $+2$ on the $\mathrm{\{Co(H_2O)_5\}}$ moiety and $\sim\!0$ on CO$_2$ along the entire path; no charge-redistribution event is observed. Complementary for the charge $+3$ results, the spin state gaps at the equilibrium point $\Delta E_{\mathrm{doublet-quartet}}$ for charge $+2$ are reported in Table~\ref{tab:eq_gaps_gas_charge_2}. 
SQD again reproduces the UCCSD(T)/CCSD(T) curve with differences up to 1.20~m$E_h$ for doublet and 0.22~m$E_h$ for quartet. The HCI curve differs from SQD energies by no more than 1.87~m$E_h$ for doublet and 0.51~m$E_h$ for quartet, at every geometry. Two consequences follow. First, the absence of an analogous repulsive feature in the charge-$+2$ states confirms that the bump on the charge-$+3$ quintet curve is not a numerical artefact of SQD post-processing or AVAS construction. Second, the diabatic crossing in Eq.~\eqref{eq:CT-diabats} is specifically enabled by the combination of (i)the higher Co oxidation state ($\mathrm{Co(III)}$, which makes $\mathrm{Co(II)}\!+\!\mathrm{CO_2^+}$ energetically accessible) and (ii) the high-spin manifold (which makes the transfer spin-allowed): the analogous internal transfer in $\mathrm{Co(II)}$ would produce $\mathrm{Co^{I}}$ and is energetically inaccessible at these distances. The control therefore supports the above mentioned charge-transfer interpretation (Subsec. Charge-$+3$ quintet dissociation curve).

\begin{table}[H]
    \caption{Doublet-quartet energy gaps $\Delta E = E(\mathrm{doublet}) - E(\mathrm{quartet})$ at equilibrium as a function of active-space size $(n_{\textrm{elec}}, n_{\textrm{mo}})$ in the gas phase for charge=2. Results from ROHF, UCCSD(T), HCI, and SQD are compared, highlighting the dependence of correlation treatment on the chosen active space.}
    \label{tab:eq_gaps_gas_charge_2}
    \centering
    \begin{tabular}{|l|l|l|l|l|}
        \hline
         $\Delta E_{\textrm{ROHF}}\textrm{/mHa}$ & ($n_{\textrm{elec}}, n_{\textrm{mo}}$) & $\Delta E_{\textrm{CCSD(T)}}\textrm{/mHa}$ & $\Delta E_{\textrm{HCI}}\textrm{/mHa}$ & $\Delta E_{\textrm{SQD}}\textrm{/mHa}$ \\ \hline
         57.8 & (23, 17) & 48.2 & 48.1 & 48.7 \\ \hline
    \end{tabular}
\end{table}

\begin{figure}
    \centering
    \includegraphics[width=0.5\linewidth]{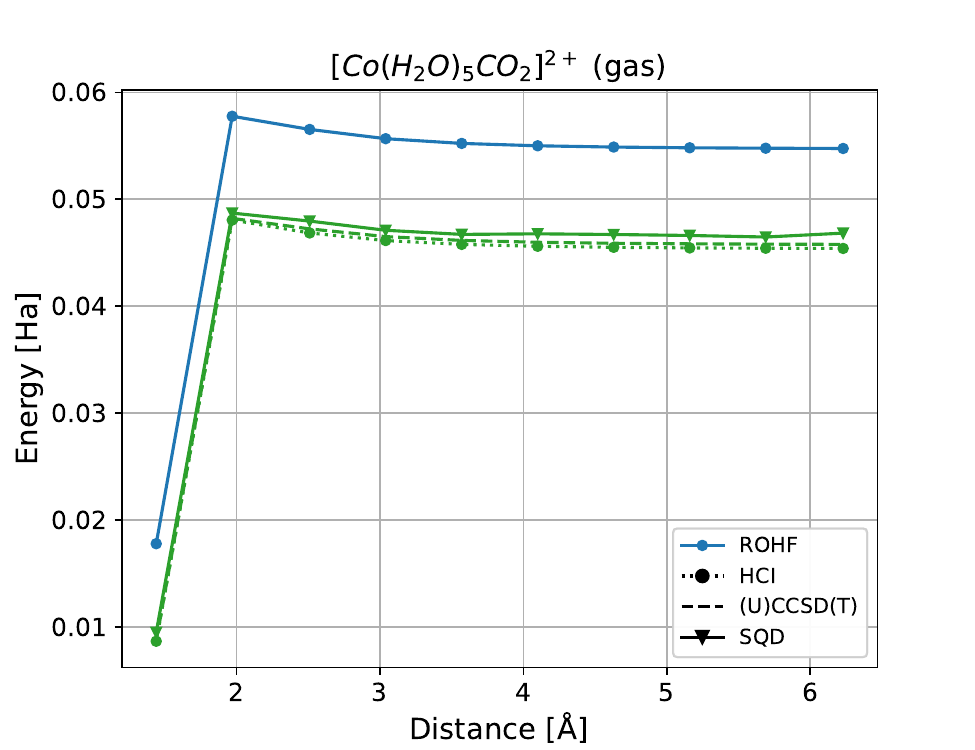}
    \caption{Quartet-doublet energy differences $\Delta E = E(\mathrm{doublet}) - E(\mathrm{quartet})$ for [Co(H$_2$O)$_5$CO$_2$]$^{2+}$ in the gas phase as a function of  $\mathrm{Co-O_{(CO_2)}}$ distance. Results are shown for increasing active-space sizes and compared across ROHF, HCI, UCCSD(T), and SQD methods.}
    \label{fig:gas_charge_2_diff_results}
\end{figure}

\begin{figure}[h]
    \centering
    \includegraphics[width=0.5\linewidth]{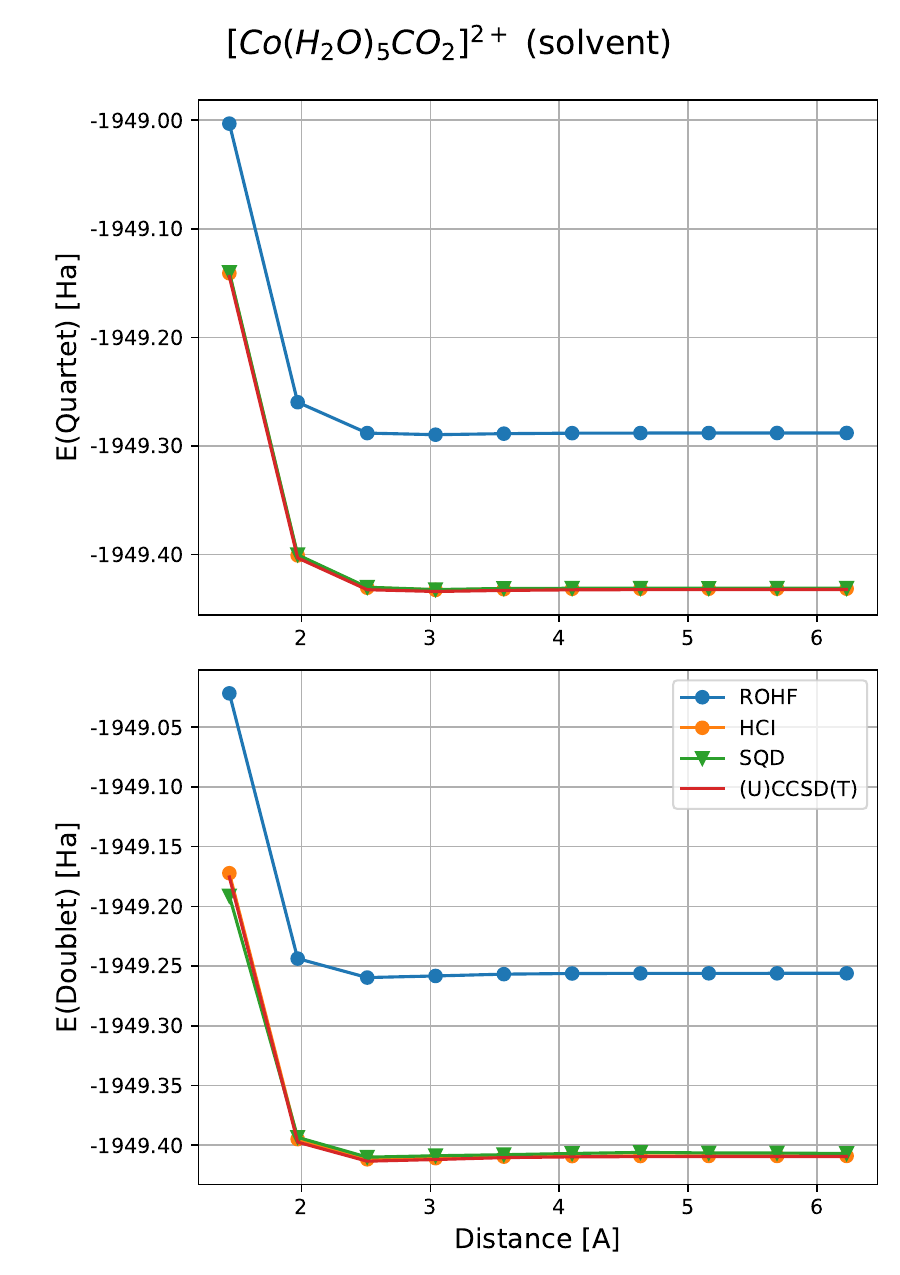}
    \caption{Figure shows potential energy curves for doublet and quartet for charge=2 for 34 qubits for solvent phase. Active space includes $(17o,12n_\alpha,11n_\beta)$ for doublet and  $(17o,13n_\alpha,10n_\beta)$ for quartet.}
    \label{fig:solvent_charge_2_results}
\end{figure}

\begin{figure}
    \centering
    \includegraphics[width=0.5\linewidth]{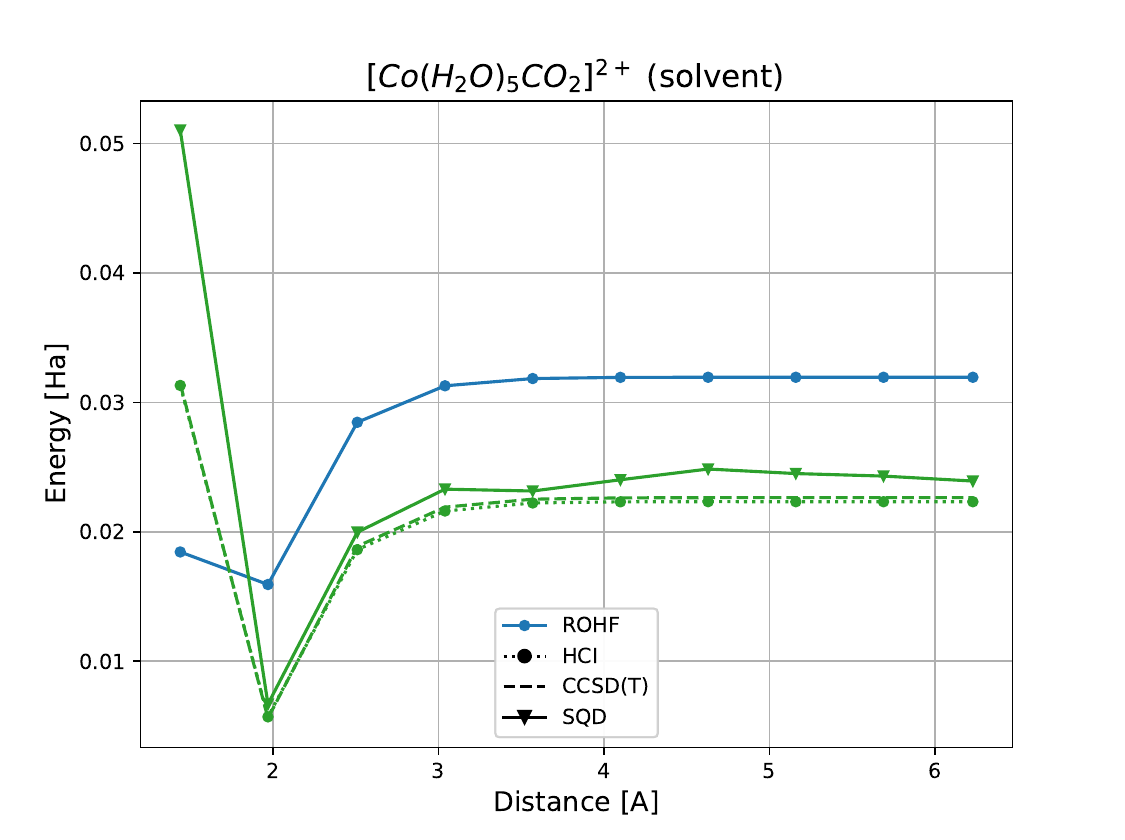}
    \caption{Quartet-doublet energy differences $\Delta E = E(\mathrm{doublet}) - E(\mathrm{quartet})$ for [Co(H$_2$O)$_5$CO$_2$]$^{2+}$ in the solvent phase as a function of  $\mathrm{Co-O_{(CO_2)}}$ distance. Results are shown for increasing active-space sizes and compared across ROHF, HCI, UCCSD(T), and SQD methods.}
    \label{fig:solvent_charge_2_diff_results}
\end{figure}

\FloatBarrier

\subsection{Implicit solvent effects}
\label{sec:solvent-results}

\subsubsection{Spin state energetics at equilibrium in implicit solvent.}
We next embed the solute in an IEF-PCM dielectric continuum at the water dielectric ($\varepsilon_r=78.36$) and repeat the equilibrium calculations for both spin states across all four active spaces (see Table~\ref{tab:active-spaces}) for $[\mathrm{Co(H_2O)_5CO_2}]^{3+}$. ROHF/IEF-PCM provides the reference determinant for the LUCJ ansatz, and the LUCJ parameters are obtained from a classical CCSD/UCCSD calculation in the same solvated active space (see Methods). Bitstrings are sampled once on \texttt{ibm\_pittsburgh}. The SCRF outer loop then updates the active-space one-electron integrals from the SQD density entirely in classical post-processing, with no additional hardware sampling. The active-space one-electron integrals are updated at each SCRF cycle from the back-rotated SQD AO-basis density $\gamma_{\mathrm{AO}}^{(k-1)}$ via Eq.~\eqref{eq:hcore-update-main}. The free-energy functional is evaluated with the standard double-counting correction $G^{(b),k}\!=\!E^{(b),k}\!+\!E_{\mathrm{nuc}}\!+\!G_{\mathrm{solv}}^{(b),k}\! \!\tfrac12\mathrm{Tr}[V^{\mathrm{solv},(k-1)}\gamma^{(b),k}]$. The SCRF loop converges within 6~iterations (typical inter-iteration $\Delta G\!<\!10^{-3}$~Ha) for all spin states at the equilibrium geometry.

The equilibrium energetics are summarised in Table~\ref{tab:eq_gaps_solvent_charge_3}. Three points are worth highlighting. First, the spin state ordering is preserved: the high-spin state remains lower in energy than the low-spin state at fixed charge in IEF-PCM, both for the charge-$+3$ quintet/singlet pair, in agreement with the gas-phase ordering. Second, the magnitude of the spin state gap shifts modestly upon solvation - for 25 orbitals CCSD(T)/UCCSD(T) the gap difference is $\Delta G_{\mathrm{singlet-quintet}}^{\mathrm{IEF\!-\!PCM}}\!-\!\Delta E_{\mathrm{singlet-quintet}}^{\mathrm{gas}}\!=\!$ -1.18~kcal\,mol$^{-1}$ for charge $+3$ , indicating that the solvent stabilises both spin state systems. Third, the SQD, (U)CCSD(T)/IEF-PCM, and HCI/IEF-PCM gaps agree within 3.02~kcal\,mol$^{-1}$, demonstrating that the open-shell extension of the SQD-IEF-PCM workflow - the ROHF reference, the $S_z$-preserving S-CORE step, and the $\hat S^2$ soft constraint - carries the gas-phase methodological reliability over to the solvated regime. 

We also verify this for the charge-$+2$ quartet/doublet pair for 34 qubit active space. We again observe that the spin state ordering is preserved at equilibrium for charge $+2$ (see Table \ref{tab:eq_gaps_solvent_charge_2}). This completes the analysis for equilibrium spin state energetics under solvation.

\begin{table}[H]
    \caption{Singlet-quintet energy gaps $\Delta E = E(\mathrm{singlet}) - E(\mathrm{quintet})$ (in mHa)  at equilibrium as a function of active-space size $(n_{\textrm{elec}}, n_{\textrm{mo}})$ in the solvent phase for charge=3.}
    \label{tab:eq_gaps_solvent_charge_3}
    \centering
    \begin{tabular}{|c|l|l|l|l|}
        \hline
        $\Delta E_{\textrm{ROHF}}$ 
        & $(n_{\textrm{elec}}, n_{\textrm{mo}})$ 
        & $\Delta E_{\textrm{CCSD(T)}}$ 
        & $\Delta E_{\textrm{HCI}}$ 
        & $\Delta E_{\textrm{SQD}}$ \\ \hline
        \multirow{4}{*}{82.5}
        & (22, 17) & 72.4 &  73.7 & 74.2 \\ \cline{2-5}
        & (24, 19) & 74.9 &  76.0 & 76.4 \\ \cline{2-5}
        & (28, 23) & 56.0 &  54.3 & 53.3 \\ \cline{2-5}
        & (30, 25) & 48.0 &  45.7 & 50.5 \\ \hline
    \end{tabular}
\end{table}

\begin{table}[H]
    \caption{Doublet-quartet energy gaps $\Delta E = E(\mathrm{doublet}) - E(\mathrm{quartet})$ at equilibrium as a function of active-space size $(n_{\textrm{elec}}, n_{\textrm{mo}})$ for solvent phase for charge=2. Results from ROHF, UCCSD(T), HCI, and SQD are compared, highlighting the dependence of correlation treatment on the chosen active space.}
    \label{tab:eq_gaps_solvent_charge_2}
    \centering
    \begin{tabular}{|l|l|l|l|l|}
        \hline
         $\Delta E_{\textrm{ROHF}}\textrm{/mHa}$ & ($n_{\textrm{elec}}, n_{\textrm{mo}}$) & $\Delta E_{\textrm{CCSD(T)}}\textrm{/mHa}$ & $\Delta E_{\textrm{HCI}}\textrm{/mHa}$ & $\Delta E_{\textrm{SQD}}\textrm{/mHa}$ \\ \hline
         15.9& (23, 17) & 5.6 & 5.7 & 6.7 \\ \hline
    \end{tabular}
\end{table}

\FloatBarrier

\subsubsection{Solvent stabilisation of the charge-$+3$ quintet dissociation.}
\begin{figure*}
    \centering
    \includegraphics[width=1\linewidth]{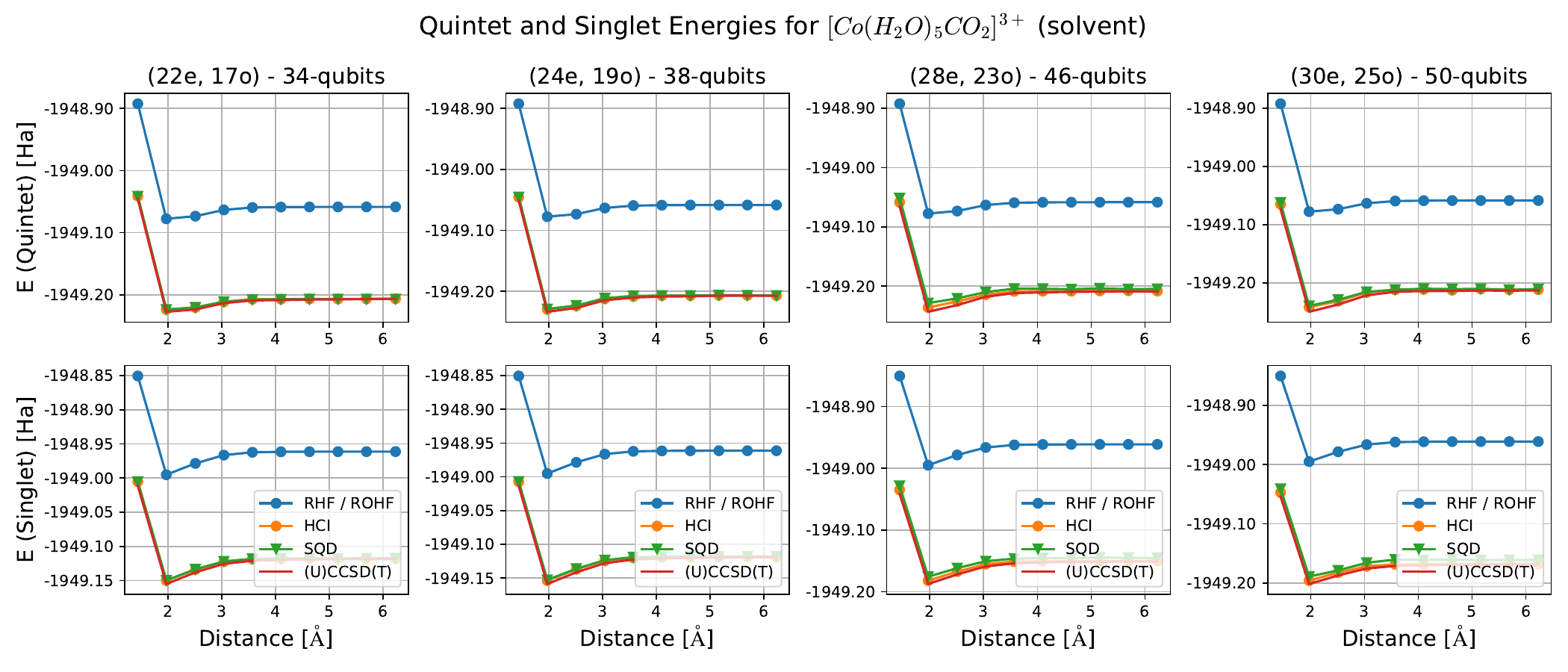}
    \caption{Potential energy curves for the quintet (top row) and singlet (bottom row) states of [Co(H$_2$O)$_5$CO$_2$]$^{3+}$ in the solvent phase for charge=3, computed for increasing active-space sizes. Results from ROHF, HCI, (U)CCSD(T), and SQD are compared as a function of the $\mathrm{Co-O_{(CO_2)}}$ distance.}
    \label{fig:solvent_charge_3_results}
\end{figure*}

The most striking solvent effect appears along the charge-$+3$ quintet dissociation coordinate (Figure~\ref{fig:solvent_charge_3_results}). Whereas the gas-phase quintet curve exhibits the pronounced repulsive feature near $r\!\approx\!3$~\AA\ (Figure~\ref{fig:gas_charge_3_results}), the IEF-PCM curve at the same active space and same spin state is monotonic and smooth: the local maximum is washed out, and the energy stabilizes gradually toward the dissociation asymptote with no anomalous structure. The charge-$+3$ singlet retains its smooth gas-phase shape under IEF-PCM, simply an offset by the solvation free energy is observed. The charge-$+2$ doublet and quartet controls (Figure~\ref{fig:solvent_charge_2_results}) show featureless curves in solvent, mirroring their gas-phase behaviour.

\begin{figure}
    \centering
    \includegraphics[width=0.5\columnwidth]{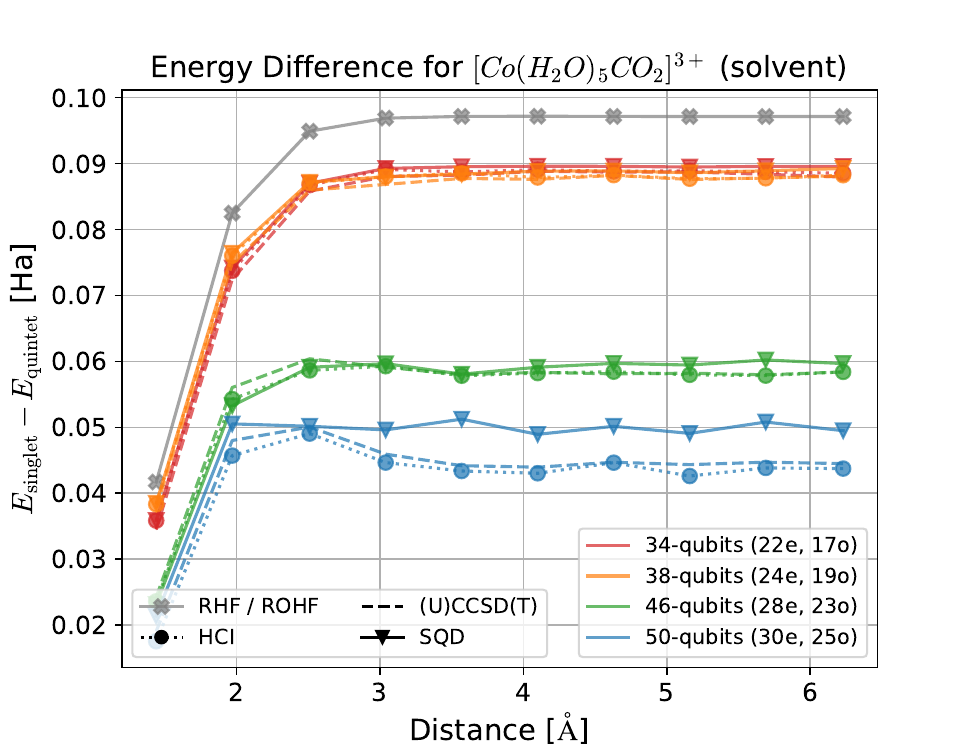}
    \caption{Singlet-quintet energy differences $\Delta E = E(\mathrm{singlet}) - E(\mathrm{quintet})$ for [Co(H$_2$O)$_5$CO$_2$]$^{3+}$ in the solvent phase as a function of $\mathrm{Co-O_{(CO_2)}}$ distance. Results are shown for increasing active-space sizes and compared across RHF/ROHF, HCI, CCSD(T)/UCCSD(T), and SQD methods.}
    \label{fig:solvent_charge_3_diff_results}
\end{figure}

\begin{figure}[h]
    \centering
    \includegraphics[width=0.5\columnwidth]{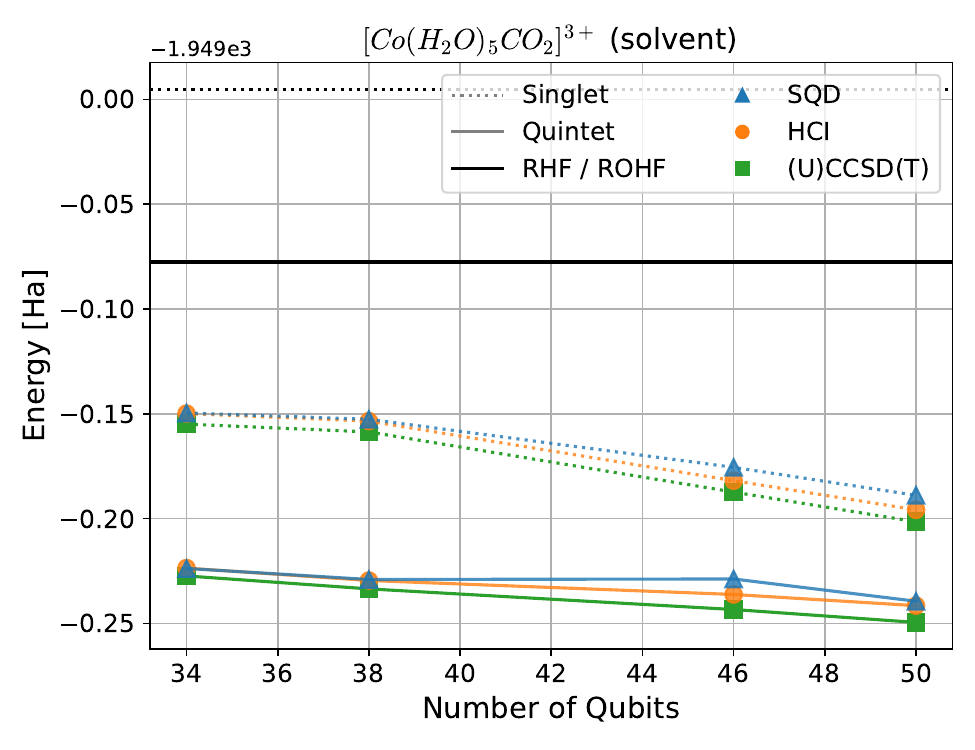}
    \caption{ % \textcolor{red}{add caption.}
    Solvent phase equilibrium energetics versus qubits: ROHF / UCCSD(T)/CCSD(T) / HCI / SQD energies at equilibrium on y-axis vs qubit count on x-axis for +3 charge singlet and quintet.}
    \label{fig:eq-vs-qubits}
\end{figure}

The mechanism is straightforward in the diabatic picture of Eq.~\eqref{eq:CT-diabats}. In the gas phase the localised 
$\mathrm{\ket{\{Co(H_2O)_5\}^{+3}~~CO_2}}$ diabat lies below the charge-separated $\mathrm{\ket{\{Co(H_2O)_5\}^{+2}~~CO^+_2}}$ diabat at short $r$ but is overtaken by it as $r$ is elongated, producing the avoided crossing. In a polar continuum, both diabats are stabilised, but to different extents: at the Born level, the solvation free energy of a single ion of charge $q$ in a cavity of radius $R$ scales as $-q^2/(2R)$, favouring the charge-localised $\mathrm{\ket{\{Co(H_2O)_5\}^{+3}~~CO_2}}$ state ($q^2\!=\!9$) over the charge-separated $\mathrm{\ket{\{Co(H_2O)_5\}^{+2}~~CO^+_2}}$ state ($q^2\!=\!4+1\!=\!5$, summed over two cavities) by an amount that grows with the dielectric constant. The differential stabilisation raises the relative energy of the charge-separated diabat, pushes the avoided crossing past the dissociation limit, and removes the local maximum along the physically accessible portion of the curve. Mulliken analysis on the back-rotated solvated 1-RDM confirms this picture: the sharp crossover seen in the gas phase is absent in solvent, and the Co-fragment charge remains close to $+3$ across the entire IEF-PCM dissociation coordinate.

With this, we establish that the open-shell SQD-IEF-PCM workflow correctly reproduces the qualitative reshaping of an open-shell, charged transition-metal potential energy surface by a continuum solvent - an environment response that is invisible to gas-phase methods, that requires the full self-consistent reaction-field machinery, and that has not previously been demonstrated within an SQD framework for an open-shell transition-metal complex. Together with the gas-phase results of Sec.\ref{sec:gas-results}, it shows that SQD is simultaneously capable of addressing (i) close-lying open-shell spin states with potential multireference character, (ii) internal electron-transfer dissociation, and (iii) environment-dependent stabilisation of charged states, at qubit counts ($\le\!50$ system qubits) that are accessible on present-generation IBM Heron R3 hardware.

\FloatBarrier
\section{Conclusions}
\label{sec:conclusions}

In this work, we show that SQD can accurately describe open-shell transition-metal chemistry across competing spin states, metal-ligand charge-transfer processes, and solvent-induced electronic reorganization.
Using the \([ \mathrm{Co(H_2O)_5 CO_2} ]^{2+/3+}\) test system, which combines multiple oxidation states, spin multiplicities, and a well-defined metal-ligand dissociation coordinate, we have shown that SQD can deliver quantitatively reliable results across a range of active-space sizes up to 50 qubits, accessible on current IBM Heron hardware. In both the gas phase and under implicit solvation, SQD reproduces CCSD(T)/UCCSD(T) and HCI benchmarks within the same active spaces, demonstrating that subspaces constructed from quantum-sampled determinants, despite hardware noise and without additional variational optimization, are sufficient to capture the underlying physical characteristics of open-shell 3d transition-metal systems.

Beyond quantitative agreement, the method captures nontrivial chemical phenomena. In particular, the gas-phase dissociation curve of the \([ \mathrm{Co(H_2O)_5 CO_2} ]^{3+}\) quintet state exhibits a pronounced nonmonotonic feature, manifested as a local maximum, which is absent in the corresponding singlet and in all charge-$+2$ states. Analysis of the one-particle density matrix reveals that this feature originates from an avoided crossing between localized and charge-separated diabatic configurations, corresponding to an internal electron-transfer process that is spin-allowed within the quintent high-spin manifold. SQD reproduces not only the energetics but also the qualitative structure of this crossover, indicating that the sampled determinant space spans both diabatic states and their coupling. The absence of analogous behavior in the charge-$+2$ doublet and quartet states provides an internal control, confirming that the observed effect is intrinsic to the interplay of oxidation state and spin symmetry rather than a methodological artifact.

Crucially, we extend the SQD framework to implicit solvation using an IEF-PCM-based self-consistent reaction-field scheme adapted to open-shell systems. This extension - requiring a ROHF reference, a $S_z$-preserving configuration recovery procedure, and a $\hat{S}^2$-constrained projected eigensolver - enables, to our knowledge, the first SQD treatment of an open-shell transition-metal complex in a dielectric environment. The solvent profoundly alters the electronic structure: the charge-transfer in the quintet dissociation curve is quenched, yielding a smooth well-behaved potential. This behavior is consistent with differential electrostatic stabilization of the competing diabatic states and demonstrates that the SQD-IEF-PCM workflow correctly captures environment-dependent electronic reorganization, a central aspect of transition-metal chemistry in realistic conditions.

These results establish SQD as a viable quantum-centric approach for transition-metal problems in which spin-state energetics, charge transfer, and environmental effects are strongly coupled. The framework developed here---combining AVAS-based active-space construction, symmetry-preserving configuration recovery, and self-consistent solvent coupling---provides a foundation for extending SQD to increasingly complex chemical settings. In future work, the present open-shell solvent framework could be combined with recent quantum-subspace developments to address molecular excited states~\cite{Barison2025}, exploit randomized Krylov-state sampling~\cite{Piccinelli2025}, and improve the construction of quantum-sampled subspace problems~\cite{Nation2026}. Building on these developments, future work can target catalytically relevant transition-metal systems, incorporate more sophisticated treatments of the chemical environment, and explore their integration with the open-shell solvent framework introduced here.

\section*{Acknowledgements}
The authors gratefully acknowledge Mario Motta for valuable discussions and insightful suggestions that helped shape the direction and development of this work.
Computational resources were provided by the Poznan Supercomputing and Networking Center (PSNC), where all calculations were performed, with support through PRACE-LAB (POIR.04.02.00-00-B001/18). The authors are particularly grateful to Adam Olszewski, Tomasz Piontek, and Bartosz Bosak for their technical support and assistance. The authors also thank Wagner Brandeburgo and Adam Kowalski (Unilever) for fruitful discussions.
Portions of the manuscript were created with Microsoft Copilot support, that was used for language refinement, organizational suggestions, and grammar and clarity checking. All scientific content, computational analyses, interpretations, and conclusions were conducted, written, reviewed and approved by the authors, who take full responsibility for the manuscript.

\section*{Supporting Information} 

Detailed methodological formulations; additional Hartree-Fock stability analyses; quantum-circuit characteristics and raw-QPU performance data; detailed SQD and HCI post-processing statistics (PDF).

\section*{Data Availability}

The data supporting the findings of this study, including molecular geometries, quantum-chemistry intermediates, and custom code, are publicly available in the \cite{kowalik_sqd} repository. The remaining software dependencies are identified in the Methods and Supporting Information.

\section*{Conflict of Interest}

The authors declare no competing financial interest.

\FloatBarrier
\bibliography{jctc_bib.bib}

% =========================================
% In general, you should place your graphics where they make logical sense; the production
% process will move them if needed.
% ========================================

\end{document}

\maketitle

\footnote{$^\$$These authors contributed equally to this work.}

% \appendix

\section{Methods in Detail}
\label{app:methods_in_detail}

%%%%%%%%%%%%%%%%%%%%%%%%%%%%%%%%%%%%%%%%%%%%%%%%%%%%%%%%%%%%
\subsection{RHF \& ROHF}
\label{app:rhf_rohf}
\begin{figure}[H]
    \centering
    \includegraphics[width=0.5\columnwidth]{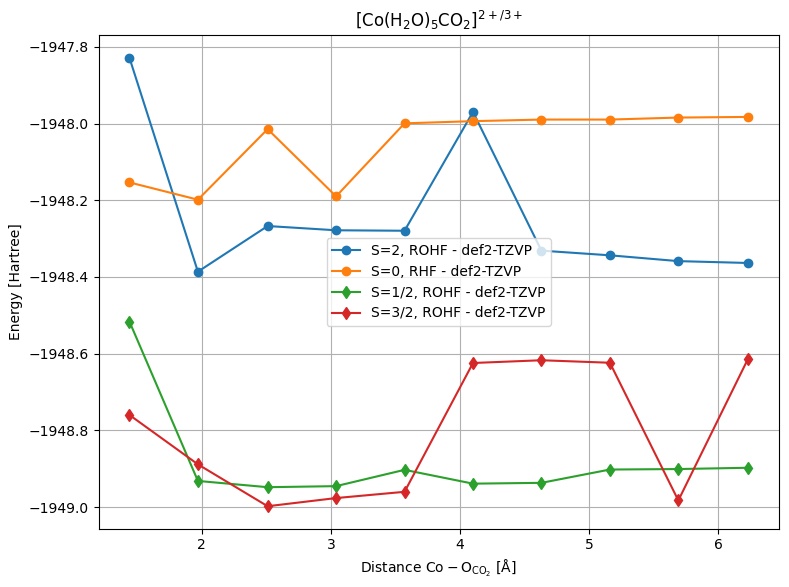}
    \caption{RHF and ROHF potential energy curves along $\mathrm{Co-O_{(CO_2)}}$  distance for Singlet, Quintet, Doublet and Quartet states before stability analysis and application of second order SCF.}
    \label{fig:rhf_rohf_before_SOSCF}
\end{figure}

For the Singlet state we perform restricted HF (RHF) calculations along the reaction path. For the Doublet, Quartet and Quintet spin states we perform ROHF calculations as a starting point. Density-Fitting\cite{Baerends1973ChemPhys,Whitten1973JCP,Dunlap1979JCP} of the 2-electron integrals, also known as the Resolution of Identity (RI) Approximation, is used in all the calculations for speed-up with corresponding auxiliary basis functions available in PySCF for the def2-tzvp basis set. In the non-relativistic treatment of the Schrödinger equation, the Hamilton operator $\hat{H}$ does not contain any reference to spin, hence it commutes with the squared total spin angular momentum operator $\hat{S}^2$ and the $\hat{S}_Z$ operator.
\begin{equation}
   [\hat{H},\hat{S}^2]=[\hat{H},\hat{S}_Z]=0
\end{equation}
Thus a proper electronic wave function should not be only an eigenfunction of $\hat{H}$ but of $\hat{S}^2$ and $\hat{S}_Z$ as well.   
The mentioned open-shell spin states correspond to configurations where all the unpaired electrons have $\alpha$-spin, hence each of the corresponding restricted open-shell determinant ($\ket{\Psi_{ROHF}^{N_\alpha}},N=1,3,4$) is an Eigenfunction of the $\hat{S}^2$ operator and trivially of the $\hat{S}_Z$ operator.\\
\begin{equation}
    \hat{S}^2 \ket{\Psi_{ROHF}^{N_\alpha}}=S(S+1)\ket{\Psi_{ROHF}^{N_\alpha}}
    \label{eq:s_square}
\end{equation}
In equation (\ref{eq:s_square}) $N_\alpha$ and $S$ represent the total number of unpaired $\alpha$-spin electrons and the total spin quantum number, respectively. Taking the $2S+1$ value leads to the mentioned multiplicities.
The RHF and ROHF calculations for all the mentioned spin states in this work show that the electronic structure is not easily described by a single determinant. First RHF and ROHF results indicated several mean-field solutions, which is visible from the corresponding energy curves, see figure \ref{fig:rhf_rohf_before_SOSCF} in the Section~\ref{app:si_rhf_rohf_before_soscf}. Subsequent stability analysis clearly shows that the corresponding RHF and ROHF wave functions are instable with respect to internal orbital rotations. Finally using the second-order SCF method implemented in PySCF alongside with one-particle reduced density matrix propagation from one point to another on the reaction path leads to smooth curves. %\phalgun
These considerations are particularly important for the present SQD workflow, where preservation of the correct spin sector and control of spin contamination directly influence the quality of the sampled determinant subspace.
%%%%%%%%%%%%%%%%%%%%%%%%%%%%%%%%%%%%%%%%%%%%%%%%%%%%%%%%%%%%
\subsection{AVAS}
\label{app:avas}
One of the challenging tasks before performing any post-HF calculation within an active space, is the selection of a proper one capturing the relevant underlying physical properties of the system correctly. Different approaches exist on how to identify the active space orbitals, ranging from chemical intuition and observation of the single-reference (HF or DFT) orbital shapes and energies to selection criteria based on (natural) orbital occupation numbers or various orbital entanglement criteria.\\ An approach justified by the fact that strong electron correlation effects stem from the presence of (near-)degenerate atomic valence orbitals with small overlap to other orbitals, is the so called atomic valence active space (AVAS) method\cite{Sayfutyarova2017}. Indeed chemical knowledge is important to decide which atomic orbitals are relevant, but no further post-HF calculation is needed to create an active space.\\
The main idea is to define a set $\{A\}$ of atomic orbitals, which are assumed to be relevant for the description of the electronic structure of the system under investigation, e.g. the 3d orbitals of a transition-metal in a complex. Then a projector $\hat{P}$ is constructed to project the molecular orbitals onto the subspace spanned by the user-defined atomic orbitals $\ket{\varphi_\mu} \in \{A\}$.
\begin{equation}
\hat{P} = \sum_{\mu,\nu \in \{A\}}\ket{\varphi_\mu}S^{-1}_{\mu \nu}\bra{\varphi_\nu}
\label{eq:avas_projector}
\end{equation}
In equation (\ref{eq:avas_projector}) the overlap matrix elements between atomic orbitals of the chosen set are defined as
\begin{equation}
    S_{\mu \nu} = \braket{\varphi_\mu|\varphi_\nu}; ~\mu, \nu \in \{A\}. 
\end{equation}
The occupied and virtual orbitals of the mean field wave function are projected separately onto the subspace defined by $\{A\}$. Hence two separate projected overlap matrices, defined by their elements
\begin{equation}
\begin{split}
    S^{occ}_{ij}&=\braket{\phi_i|\hat{P}|\phi_j}\\
    S^{vir}_{ab}&=\braket{\phi_a|\hat{P}|\phi_b},
\end{split}
\label{eq:socc_svir_avas}
\end{equation}
are built. In equation (\ref{eq:socc_svir_avas}) $\ket{\phi_i}, i \in occ$ and $\ket{\phi_a}, a \in vir$ represent occupied and virtual molecular orbitals respectively. Diagonalizing $S^{occ}$ and $S^{vir}$, defined by their elements in equation (\ref{eq:socc_svir_avas}), leads to
\begin{equation}
    \begin{split}
        S^{occ}U^{occ}&=s^{occ}_{diag}U^{occ}\\
        S^{vir}U^{vir}&=s^{vir}_{diag}U^{vir}.
    \end{split}
    \label{eq:uocc_uvir_avas}
\end{equation}
In equation (\ref{eq:uocc_uvir_avas}) $U^{occ}$ and $U^{vir}$ define unitary transformation matrices for the occupied and virtual space respectively. Each column of $U^{occ}$ and $U^{vir}$ represent an eigenvector of $S^{occ}$ and $S^{vir}$ with corresponding eigenvalue in $s^{occ}_{diag}$ and $s^{vir}_{diag}$, respectively. According to a predefined threshold ($\tau$) only column vectors of $U^{occ}$ and $U^{vir}$ are kept which are larger than or equal to $\tau$ resulting in $\tilde{U}^{occ}$ and $\tilde{U}^{vir}$. These are used to  construct MO coefficient matrices transforming from AO-basis to AVAS occupied ($\tilde{C}^{occ}$) and virtual space ($\tilde{C}^{vir}$), respectively.
\begin{equation}
    \begin{split}
        \tilde{C}^{occ} &= C^{occ}\tilde{U}^{occ}\\
        \tilde{C}^{vir} &= C^{vir}\tilde{U}^{vir}\\
    \end{split}
\end{equation}
In order to selectively adding more occupied orbitals without changing the number of virtual orbitals in the active space or vice versa, we used two separate AVAS thresholds $\tau_{occ}$ and $\tau_{vir}$.
%%%%%%%%%%%%%%%%%%%%%%%%%%%%%%%%%%%%%%%%%%%%%%%%%%%%%%%%%%%%
\subsection{CCSD}
\label{app:ccsd}
The movement of two electrons with opposite spin is not correlated in case of a single SD. One of the widely used methods to capture this dynamic correlation of electrons correctly is Coupled Cluster\cite{Cizek1969,PurvisBartlett1982,ScuseriaSchaefer1989}. The CC wave function $\ket{CC}$ is defined using an exponential ansatz as shown in the following.
\begin{equation}
    \ket{CC}=\mathrm{exp(\textbf{T})}\ket{\Psi_0}
    \label{eq:CC_wave_function}
\end{equation}
In CCSD the cluster operator $\mathrm{\textbf{T}}$ from equation (\ref{eq:CC_wave_function}) is defined as the sum of the singles and doubles cluster operator.
\begin{equation}
    \begin{split}
        \mathrm{\textbf{T}}&=\mathrm{\textbf{T}_1}+\mathrm{\textbf{T}_2}\\
        \mathrm{\textbf{T}_1}&=\sum_{IA}t^I_Aa^{\dagger}_Aa_I\\
        \mathrm{\textbf{T}_2}&=\frac{1}{4}\sum_{IJAB}t^{IJ}_{AB}a^{\dagger}_Aa^{\dagger}_Ba_Ja_I
    \end{split}
    \label{eq:single_doubles_cluster_operator}
\end{equation}
In equation (\ref{eq:single_doubles_cluster_operator}) $t^I_A$ and $t^{IJ}_{AB}$ represent elements of the singles and doubles amplitudes, respectively. The indices $I,J$ and $A,B$ represent occupied and virtual molecular spin orbitals, respectively. Whereas the creation ($a^{\dagger}_A,~a^{\dagger}_B$) and annihilation operators ($a_J,~a_I$) act explicitly on virtual and occupied orbitals of the reference determinant $\ket{\Psi_0}$ to generate singly and doubly excited determinants.
The CCSD energy is not obtained as an ordinary quantum mechanical expectation value of the electronic Hamiltonian, but rather as a similarity transformation.
\begin{equation}
    E_{CCSD} = \braket{\Psi_0|\mathrm{\mathrm{exp(-\textbf{T}_1-\textbf{T}_2})}\hat{H}\mathrm{\mathrm{exp(\textbf{T}_1+\textbf{T}_2})}|\Psi_0}
    \label{eq:ccsd_energy}
\end{equation}
The final CCSD energy is calculated using the converged amplitudes from the CC singles and doubles amplitudes equations $\Omega_{\mu_1}$ and $\Omega_{\mu_2}$.
\begin{equation}
        \Omega_{\mu_i}=\braket{\mu_i|\mathrm{\exp(-\textbf{T})}\hat{H}\mathrm{\exp(\textbf{T})}|\Psi_0}\overset{!}{=}0;~i=1,2
\end{equation}
In our CCSD calculations we restricted the singles and doubles excitations to occupied and virtual orbitals of the chosen active space, i.e. the summation over $IJAB$ in equation (\ref{eq:single_doubles_cluster_operator}) runs only over the indices belonging to the active space.
Since we do not employ spin-adapted CC equations for Doublet, Quartet and Quintet, the resulting method is the unrestricted CCSD (UCCSD).\\
To approximately include the leading contribution from the connected triple excitations to the CCSD/UCCSD energy, we additionally employed the perturbative triples correction, which is calculated from the singles and doubles amplitudes of CCSD/UCCSD, leading to the CCSD(T)/UCCSD(T) energy. Again, the triples amplitudes indices are restricted to occupied and virtual orbitals in the active space. %\phalgun
In all cases, CCSD/UCCSD calculations are restricted to the same active spaces as used in SQD and HCI, ensuring consistency of the underlying Hamiltonian and wave function.
%%%%%%%%%%%%%%%%%%%%%%%%%%%%%%%%%%%%%%%%%%%%%%%%%%%%%%%%%%%%
\subsection{HCI - gas and solvent phase}
\label{app:hci}
According to the Slater-Condon rules, Hamiltonian matrix elements between
Slater determinants vanish if the determinants differ by more than two spin
orbitals. Consequently, each determinant is directly coupled only to a small
subset of the full determinant space. The magnitudes of the nonzero
Hamiltonian couplings vary over many orders of magnitude~\cite{Holmes2016}.
This observation motivates selected configuration interaction (SCI)
methods~\cite{Huron1973,Holmes2016,Tubman2020}, which approximate the full CI
wave function by constructing a compact variational space containing only the
most important determinants.

In Heat-Bath CI (HCI)~\cite{Holmes2016}, the variational wave function is
written as
\begin{equation}
    \ket{\Psi_{\mathrm{Var}}}
    =
    \sum_{i \in \mathrm{Var}} c_i \ket{D_i},
\end{equation}
where $\mathrm{Var}$ denotes the current variational determinant space.
Starting from an initial determinant, HCI iteratively enlarges
$\mathrm{Var}$ by adding external determinants $\ket{D_a}$ that are strongly
coupled to determinants already present in $\mathrm{Var}$. The selection
criterion is
\begin{equation}
    \max_{i \in \mathrm{Var}} |H_{ai} c_i| > \epsilon_1 ,
    \label{eq:hci_selection}
\end{equation}
where $H_{ai}=\braket{D_a|\hat{H}|D_i}$ and $c_i$ is the CI coefficient of
determinant $\ket{D_i}$ in the current variational wave function. The threshold
$\epsilon_1$ controls the size of the variational space: smaller values of
$\epsilon_1$ lead to larger determinant spaces and systematically approach the
FCI limit within the chosen active space.

After convergence of the variational HCI calculation, the missing contribution
from determinants outside $\mathrm{Var}$ can be estimated using a second-order
Epstein-Nesbet perturbative correction,
\begin{equation}
    \Delta E_2
    =
    \sum_{a \notin \mathrm{Var}}
    \frac{
    \left(
    \sum_{i \in \mathrm{Var}} H_{ai} c_i
    \right)^2
    }{
    E_{\mathrm{Var}} - H_{aa}
    },
    \label{eq:enpt2}
\end{equation}
where $E_{\mathrm{Var}}$ is the variational HCI energy and $H_{aa}$ is the
diagonal Hamiltonian matrix element of the external determinant $\ket{D_a}$.
In practice, the perturbative correction is evaluated with a second threshold
$\epsilon_2$, which restricts the sums to the most important couplings and
thereby reduces the cost of the perturbative step. The final active-space HCI
energy is obtained as
\begin{equation}
    E_{\mathrm{act}}^{\mathrm{HCI}}
    =
    E_{\mathrm{Var}} + \Delta E_2 .
    \label{eq:shci_energy}
\end{equation}

In the gas phase, HCI was applied to a fixed active-space Hamiltonian obtained
after integrating out the inactive doubly occupied core orbitals. The inactive
core contributes both a scalar energy, $E_{\mathrm{core}}$, and a
Coulomb-exchange potential to the active-space one-electron integrals. The
active-space Hamiltonian solved by HCI can be written in second-quantized form
as
\begin{equation}
    \begin{split}
        \hat{H}_{\mathrm{act}}^{\mathrm{gas}} = &\sum_{PQ \in \mathrm{act}} h^{\mathrm{eff}}_{PQ} \hat{a}^{\dagger}_{P} \hat{a}_{Q} 
    +
    \frac{1}{2} \sum_{PQRS \in \mathrm{act}}(PQ|RS)
    \hat{a}^{\dagger}_{P}
    \hat{a}^{\dagger}_{R}
    \hat{a}_{S}
    \hat{a}_{Q}.
    \end{split}
    \label{eq:hci_active_hamiltonian}
\end{equation}
Here, $\hat{a}^{\dagger}_{P}$ and $\hat{a}_{P}$ create and annihilate an
electron in spin orbital $P$, respectively. The capital indices
$P,Q,R,S$ run over active spin orbitals. The quantities
$h^{\mathrm{eff}}_{PQ}$ are the effective one-electron integrals in the
active spin-orbital basis, and $(PQ|RS)$ denotes the corresponding
two-electron repulsion integrals in chemist's notation. The effective
one-electron integrals contain the one-electron core Hamiltonian and the
mean-field Coulomb-exchange interaction with the frozen inactive core
orbitals.

The total gas-phase HCI energy is then obtained by adding the scalar
inactive-core contribution,
\begin{equation}
    E_{\mathrm{HCI}}^{\mathrm{gas}}
    =
    E_{\mathrm{act}}^{\mathrm{HCI}}
    +
    E_{\mathrm{core}} .
    \label{eq:hci_gas_energy}
\end{equation}

For solvent calculations, HCI was embedded in a self-consistent continuum
reaction-field cycle. The continuum solvent modifies only the one-electron
part of the Hamiltonian through a density-dependent reaction-field potential
$v_{\mathrm{RF}}[\rho]$, while the active-space two-electron integrals remain
unchanged. Thus, in reaction-field iteration $k$, the active-space Hamiltonian
becomes
\begin{equation}
    \begin{split}
        \hat{H}_{\mathrm{act}}^{(k)}=&\sum_{PQ \in \mathrm{act}}h_{PQ}^{\mathrm{eff},(k)} \hat{a}^{\dagger}_{P} \hat{a}_{Q} +
    \frac{1}{2}\sum_{PQRS \in \mathrm{act}}(PQ|RS)
    \hat{a}^{\dagger}_{P}
    \hat{a}^{\dagger}_{R}
    \hat{a}_{S}
    \hat{a}_{Q},
    \end{split} 
    \label{eq:hci_pcm_hamiltonian}
\end{equation}
with
\begin{equation}
    h_{PQ}^{\mathrm{eff},(k)}
    =
    \left\langle
    P
    \left|
    \hat{h}_{\mathrm{core}}^{\mathrm{gas}}
    +
    \hat{V}_{\mathrm{core}}
    +
    \hat{v}_{\mathrm{RF}}[\rho^{(k-1)}]
    \right|
    Q
    \right\rangle .
    \label{eq:hci_pcm_h1eff}
\end{equation}
Here, $\hat{V}_{\mathrm{core}}$ denotes the Coulomb-exchange potential generated by the inactive doubly occupied core orbitals. In the first reaction-field iteration, no HCI-generated solvent potential is available yet; therefore, the initial HCI density is obtained without the additional $v_{\mathrm{RF}}[\rho]$ contribution. This density is then transformed back to the AO basis, combined with the inactive-core density, and used to compute the first solvent reaction field. Subsequent HCI calculations use the reaction-field potential generated from the previous HCI density, and the cycle is repeated until the correlated HCI density and the solvent response are mutually consistent.

%%%%%%%%%%%%%%%%%%%%%%%%%%%%%%%%%%%%%%%%%%%%%%%%%%%%%%%%%%%%
\subsection{Sample-Based Quantum Diagonalization} 
\label{app:sqd}

SQD\cite{RobledoMoreno2025} is a hybrid quantum-classical method for the lowest eigenstate of a molecular Hamiltonian projected into an active space. SQD builds upon quantum-selected Configuration Interactions idea, effectively a Selected Configuration Interaction method~\cite{Kanno2023}, that utilizes quantum computers, to choose the subspace to project the Hamiltonian of the system to, and then find the accurate enough the multi-configuration ground-state (or excited state) and its respective energy within this subspace. What distinguishes SQD is the source of the subspace - determinants are drawn from measurements of an approximate ground state prepared on a quantum processor, with the device acting as a sampling oracle rather than a part of a variational engine. This makes SQD substantially more robust to hardware noise than VQE and far less depth-demanding than quantum phase estimation. We use the LUCJ ansatz\cite{Motta2023} for state preparation,
\begin{equation}
  \lvert \Phi_{\mathrm{qc}} \rangle
  = e^{\hat{K}_{2}}\,e^{i\hat{J}_{1}}\,e^{\hat{K}_{1}}\,
    \lvert \mathbf{x}_{\mathrm{ref}} \rangle ,
  \label{eq:lucj}
\end{equation}
where $\hat{K}_{1}, \hat{K}_{2}$ are one-body orbital rotations, $\hat{J}_{1}$ is a density-density Jastrow factor, and $\lvert \mathbf{x}_{\mathrm{ref}}\rangle$ is the Hartree-Fock reference determinant under the Jordan-Wigner mapping. LUCJ parameters are fixed without optimization from the $t_{1}, t_{2}$ amplitudes of a classical UCCSD/CCSD calculation on the same active space via double factorization\cite{RobledoMoreno2025,Kaliakin2025}, so the ansatz is a shallow, hardware-efficient truncation of the classical coupled-cluster wavefunction whose two-qubit gate depth is essentially independent of system size on heavy-hex topologies.

Bitstrings $\tilde{\chi}$ measured in the computational basis are corrupted by hardware noise that violates total particle number $\hat N$ and spin-$z$ projection $\hat S_z$. The self-consistent configuration recovery (S-CORE) loop\cite{RobledoMoreno2025} restores both symmetries by probabilistically flipping bits in each measurement outcome toward the current orbital-occupancy distribution $n_{p\sigma} = \langle\hat a^{\dagger}_{p\sigma}\hat a_{p\sigma}\rangle$, producing a recovered configuration pool $\chi_R$. The recovered pool is partitioned into $K$ random batches, the active-space Hamiltonian is projected onto the span of each batch and diagonalized via Davidson iteration to yield $(E^{(b)}, \lvert\psi^{(b)}\rangle)$, and updated occupancies are extracted from the batch-averaged spin-summed one-particle reduced density matrix
\begin{equation}
  \gamma_{pq}
  = \frac{1}{K}\sum_{b=1}^{K}\!\!
    \langle\psi^{(b)}\rvert\,
      \hat{a}^{\dagger}_{p\alpha}\hat{a}_{q\alpha}
      + \hat{a}^{\dagger}_{p\beta}\hat{a}_{q\beta}
    \,\lvert\psi^{(b)}\rangle .
  \label{eq:1rdm}
\end{equation}
The diagonal of $\gamma$ feeds back into S-CORE; the active-space Hamiltonian is held fixed throughout. The loop terminates when the inter-iteration energy change falls below a set threshold ($10^{-6}$~Ha in our calculations) or the occupancies stabilize, and the reported energy is $E_{\mathrm{SQD}} = \min_{b} E^{(b)} + E_{\mathrm{nuc}}$.

In the present work, we apply the open-shell SQD formulation introduced in Ref.~\cite{Liepuoniute2025}. For states with $N_\alpha \neq N_\beta$, configuration recovery is performed separately for the $\alpha$- and $\beta$-spin registers so that each recovered determinant has the target electron numbers $(N_\alpha,N_\beta)$ and therefore remains in the desired particle-number and $S_z$ sector. The spin-inversion symmetrization used in closed-shell SQD calculations is disabled (\texttt{symmetrize\_spin=False}), because exchanging the $\alpha$- and $\beta$-spin occupations would move configurations outside the target $S_z$ sector. Apart from this modification, the sampling, self-consistent configuration-recovery, batching, and projected diagonalization procedures follow the standard SQD workflow. The coupling of this open-shell SQD treatment to an implicit-solvent environment is described in the following subsections.

%%%%%%%%%%%%%%%%%%%%%%%%%%%%%%%%%%%%%%%%%%%%%%%%%%%%%%%%%%%%
\subsection{IEF-PCM for Open-Shell Solutes}
\label{app:iefpcm}

Implicit solvation replaces the explicit solvent with a polarizable dielectric continuum, with the solute occupying a molecule-shaped cavity. The solute-solvent interaction enters as a perturbation $\hat V_{\mathrm{int}}$ that augments the in vacuo Hamiltonian $\hat H^{0}$ to give $\hat H^{0} + \hat V_{\mathrm{int}}$. Because $\hat V_{\mathrm{int}}$ is itself a functional of the solute density, the eigenvalue problem is nonlinear and must be solved self-consistently - the self-consistent reaction field (SCRF) condition. In the integral equation formalism (IEF) of PCM, the three-dimensional Poisson problem in the dielectric is recast as a boundary-element problem on the cavity surface, yielding an apparent surface charge $\sigma(s)$ obtained directly from the solute electrostatic potential $\varphi^{\rho}(s)$~\cite{Tomasi2005,Mennucci1997,Cances1997}. The solvent contribution to the Fock-like one-electron operator is
\begin{equation}
  V^{\mathrm{solv}}_{\mu\nu}
  = \int_{\Gamma} \mathrm{d}s\,\sigma(s)\,\phi_{\mu\nu}(s) ,
  \label{eq:vsolv-fock}
\end{equation}
with $\phi_{\mu\nu}(s)$ the electrostatic potential at $s$ generated by the basis-function pair $\phi_{\mu}\phi_{\nu}$. This $V^{\mathrm{solv}}$ realizes $\hat V_{\mathrm{int}}$ in the electronic structure problem.

For an open-shell solute, two aspects deserve explicit comment. First, the IEF-PCM surface charge is driven by the \emph{total} electronic density $\rho = \rho_\alpha + \rho_\beta$, not the spin density: the continuum is, to the order at which IEF-PCM is formulated, spin-agnostic, and the unpaired-electron character enters the solvation response only indirectly through how open-shell correlation reshapes $\rho(\mathbf{r})$. Second, the reference determinant used to build $\hat H^{0}$ must itself be consistent with the target spin state. We take the restricted open-shell Hartree-Fock (ROHF) solution in the presence of IEF-PCM as the reference, so that $\hat H^{0}$ retains $\hat S_z$ and $\hat S^{2}$ as good symmetries and the spin contamination that complicates UHF-based PCM treatments is avoided.

As is standard in post-Hartree-Fock IEF-PCM implementations~\cite{Caricato2012,Cammi2009,Caricato2018,Castaldo2022}, we adopt the frozen-reaction-field approximation: the solvent response is determined by the correlated one-body density, while the two-electron integrals remain the bare in vacuo integrals. This decouples the expensive two-electron solvent response from the correlated eigenvalue problem while preserving the dominant electrostatic coupling. Nonelectrostatic contributions to the free energy of solvation $G_{\mathrm{solv}}$ - cavitation, dispersion, and Pauli repulsion - are not captured by IEF-PCM and would require an SM\textit{x}-type treatment~\cite{Chambers1996,Marenich2009}; their inclusion is deferred to future work.

%%%%%%%%%%%%%%%%%%%%%%%%%%%%%%%%%%%%%%%%%%%%%%%%%%%%%%%%%%%%
\subsection{Open-Shell SQD with IEF-PCM}
\label{app:sqd-iefpcm}

Three modifications extend the closed-shell SQD-IEF-PCM workflow of Kaliakin \textit{et al.}\cite{Kaliakin2025} to arbitrary $(N_{\alpha}, N_{\beta})$ active spaces: an ROHF reference determinant, an $S_z$-preserving S-CORE step, and an $\hat S^{2}$ constraint on the projected eigenproblem. The inner SQD loop of the main text Section~2.6 is then wrapped inside an outer SCRF iteration that updates the active-space integrals from the SQD density at each cycle.

\paragraph{Reference determinant and symmetry sectors.}
For an $(N_{\alpha}, N_{\beta}; N_{\mathrm{orb}})$ active space, the LUCJ reference (eq.~\ref{eq:lucj}) is the ROHF determinant obtained from an ROHF IEF-PCM calculation on the target molecule in the solvent of interest, and LUCJ parameters come from a classical ROHF-based UCCSD/CCSD calculation on the same solvated active space. We follow Liepuoniute \textit{et al.}\cite{Liepuoniute2025} in assigning spin-up and spin-down orbitals to disjoint qubit registers connected by auxiliary qubits that mediate the density-density Jastrow couplings.

\paragraph{S-CORE for $S_z \neq 0$.}
S-CORE proceeds as in main text Section~2.6, with two adjustments. The spin-inversion symmetrization that closed-shell singlets use to enforce $\langle \hat S_z \rangle = 0$ is disabled ({\ttfamily symmetrize\_spin = False}); for $N_\alpha \neq N_\beta$ that operation would move configurations out of the physical $S_z$ sector. Because open-shell occupation number vectors are not always eigenfunctions of $\hat S^{2}$, the sampled subspace is not automatically an $\hat S^{2}$ eigenspace, and we follow Robledo-Moreno \textit{et al.}\cite{RobledoMoreno2025} in imposing the exact $\hat S^{2}$ eigenvalue $S(S+1)$ as a soft constraint ({\ttfamily spin\_sq}) in the PySCF selected-CI eigensolver, mitigating residual spin contamination.

\paragraph{SCRF outer loop.}
Whereas the gas-phase SQD loop of main text Section~2.6 updates only the occupancies that guide S-CORE while holding the Hamiltonian fixed, the SCRF condition requires the Hamiltonian itself to be rebuilt from the SQD density at each outer cycle. At SCRF iteration $k$, the active-space one-electron integrals are updated as
\begin{equation}
  h_{pq}^{(k)}
  = h_{pq}^{(0)} + V^{\mathrm{solv}}_{pq}\!\left[\gamma^{(k-1)}\right] ,
  \label{eq:hcore-update}
\end{equation}
where $h_{pq}^{(0)}$ are the in vacuo MO-basis one-electron integrals, $\gamma^{(k-1)}$ is the AO-basis total 1-RDM obtained by back-rotation of eq.~\ref{eq:1rdm}, and $V^{\mathrm{solv}}[\gamma]$ is the IEF-PCM Fock matrix of eq.~\ref{eq:vsolv-fock}. The two-electron integrals remain at their in vacuo values. The total free energy at iteration $k$ for batch $b$ is
\begin{equation}
  G^{(b),\,k}
  = E^{(b),\,k} + E_{\mathrm{nuc}} + G_{\mathrm{solv}}^{(b),\,k}
  - \tfrac{1}{2}\,\mathrm{Tr}\!\left[
      V^{\mathrm{solv},\,(k-1)}\,\gamma^{(b),\,k}
    \right] ,
  \label{eq:gibbs-per-batch}
\end{equation}
the last term being the standard double-counting correction for polarization potentials carried over between cycles. Bitstrings are sampled on the QPU only \emph{once}; the SCRF iteration is a purely classical post-processing loop. This relies on the assumption that an LUCJ ansatz produces samples whose coverage of the important determinants is sufficient to span the solvated wavefunction - an assumption used previously in closed-shell SQD-IEF-PCM\cite{Kaliakin2025} and VQE IEF-PCM\cite{Castaldo2022}. In this work, we have demonstrated the validity of this assumption in the context of an open-shell charged complex.

\paragraph{Output quantities.}
The free energy of solvation is
\begin{equation}
  \Delta G_{\mathrm{solv}}
  = G_{\mathrm{solvated}} - E_{\mathrm{gas}} ,
  \label{eq:dgsolv}
\end{equation}
with $G_{\mathrm{solvated}}$ the final SQD-IEF-PCM free energy (minimum over batches of $G^{(b),\,k_{\max}}$) and $E_{\mathrm{gas}}$ the gas-phase SQD energy at the same geometry, active space, basis, and sampling protocol. Using matched SQD energies on both sides cancels method-dependent correlation errors to a large extent, isolating the solvation contribution. Schematic workflows for the gas-phase and solvated pipelines are shown in the main text Figure~1~in the middle panel.
\FloatBarrier

%%%%%%%%%%%%%%%%%%%%%%%%%%%%%%%%%%%%%%%%%%%%%%%%%%%%%%%%%%%
\section{RHF and ROHF results before stability analysis and second order SCF}
\label{app:si_rhf_rohf_before_soscf}

\FloatBarrier

\section{Detailed quantum circuit and raw QPU results analysis}
\label{app:si_detailed_quantum_circuit_results}

This section goes through the details of circuits run on QPU, to obtain the SQD determinants input, and also the raw QPU results quality of those circuits. In particular it compares the results from:
\begin{itemize}
    \item the circuits transpiled and run using the readily available tools, called as "default"
    \item the runs using the Q-CTRL tool Fire Opal\cite{qctrl_fireopal} that includes transpilation, QPU runs workflow handling also applying error suppression
\end{itemize}

Circuit depth and number of gates (also in the 2-qubit gates only variation) is a starting point of our analysis, since it's a critical metric for neat-term quantum computations, with the direct impact on the susceptibility of circuits to gate errors, crosstalk and decoherence. 

For each medium (gas, solvent), spin state and the active space considered for our system, we constructed quantum circuits corresponding to the electronic structure simulations, as described in the main text. The logical circuits were then transpiled using 
\begin{itemize}
    \item \textbf{Default case:} Qiskit's 2.2.1 transpiler with optimization level 3, the initial layout following the 'zig-zag' pattern mapping LUCJ ansatz into heavy-hex IBM's Heron architecture. The initial layout was found using rustworkx library and custom scoring finding the mapping that is the least erroneous. The code can be found in the references from data availability section. Transpiler's pass manager's "pre\_init" stage was extended with with `ffsim.qiskit` (version 0.0.59) PRE\_INIT pass.
    \item \textbf{Q-CTRL's Fire Opal case:} Fire Opal's included transpiler, that takes previous workflow transpiled circuits as an input. 
\end{itemize}

All other aspects of the circuit construction, including gate sets, target connectivity, and compilation constrains, were kept identical across both approaches, to enable a direct comparison. No circuit split or recomposition with e.g. tensor-network approximations, was applied.

\begin{figure}
    \centering
    \text{A) Gas phase}
    \includegraphics[width=0.95\linewidth]{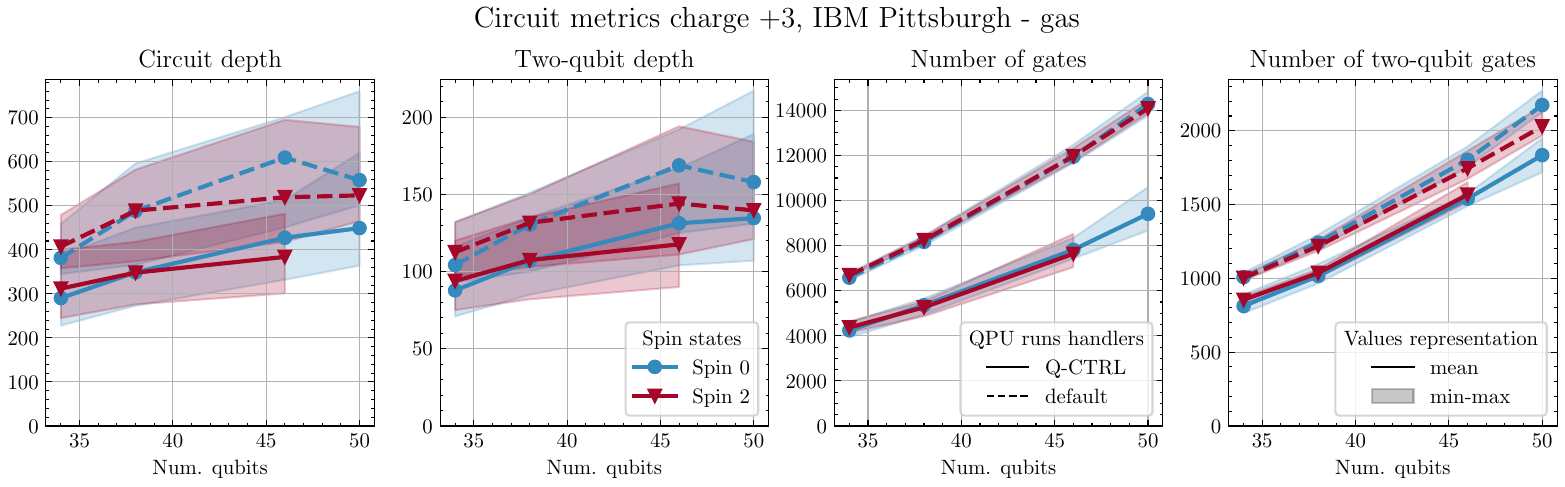}
    \vspace{\baselineskip}
    \text{B) Implicit solvent}
    \includegraphics[width=0.95\linewidth]{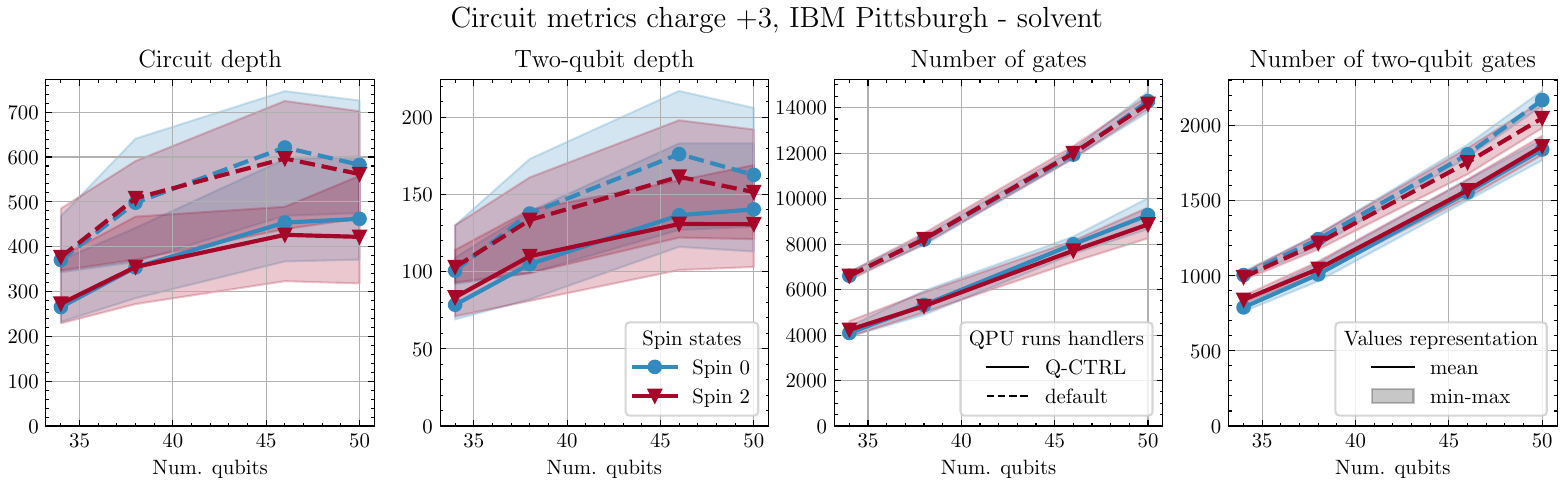}

    \caption{Circuit metrics as a function of qubit number for the charge $+3$ system in (A) the gas phase and (B) implicit solvent. The analysis was performed on the exact circuits executed on \textsf{ibm\_pittsburgh} to generate the inputs for the SQD calculations reported in this work. Circuit depth, two-qubit depth, total gate count, and two-qubit gate count are shown for spin states 0 and 2, comparing default transpilation (dashed lines) with Q-CTRL's Fire Opal (solid lines). Markers denote the corresponding means, while shaded regions indicate the minimum-to-maximum ranges. Results for the largest active space of the quintet state in the gas phase are unavailable because changes to the QPU-results retrieval handlers prevented their recovery.}
    \label{fig:SI_qc_metrics_charge_3}
\end{figure}

\begin{figure}
    \centering
    \text{A) Gas phase}
    \includegraphics[width=0.95\linewidth]{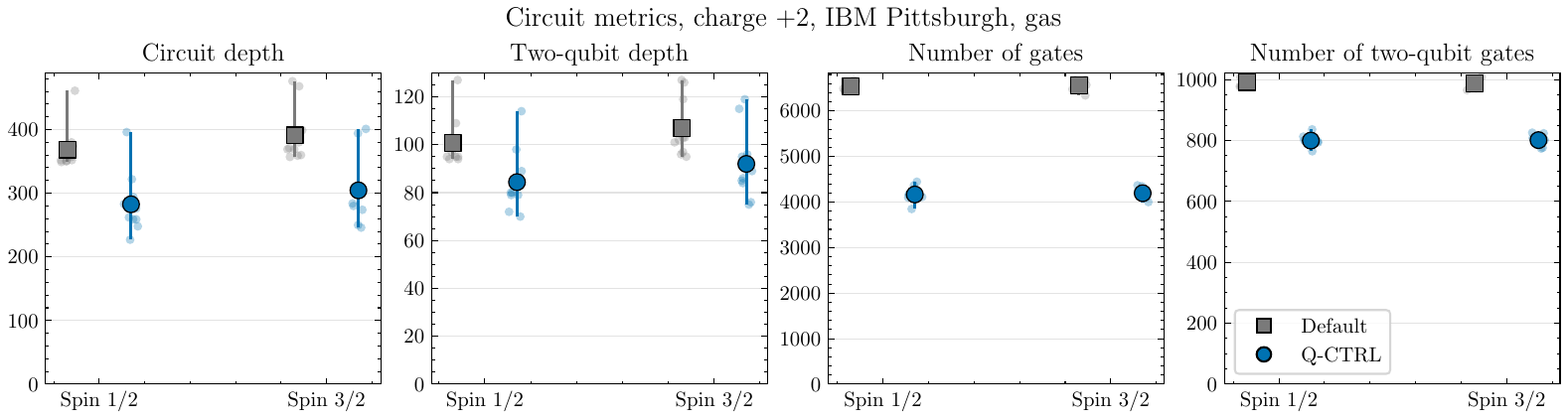}
    \vspace{\baselineskip}
    \text{B) Implicit solvent}
    \includegraphics[width=0.95\linewidth]{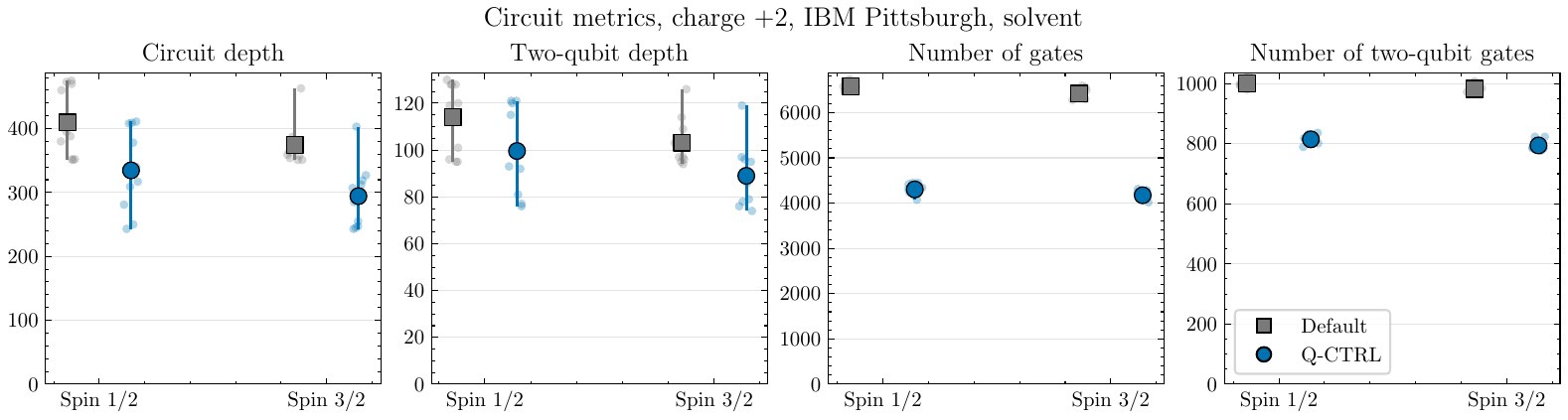}

    \caption{Circuit metrics for the fixed 34-qubit active space of the charge $+2$ system in (A) the gas phase and (B) implicit solvent. The analysis was performed on the exact circuits executed on \textsf{ibm\_pittsburgh} to generate the inputs for the SQD calculations reported in this work. Circuit depth, two-qubit depth, total gate count, and two-qubit gate count are shown for doublet and quartet spin states, comparing default transpilation and Q-CTRL's Fire Opal. Small points represent individual circuits, large markers denote the corresponding means, and vertical lines indicate the minimum-to-maximum ranges.}
    \label{fig:SI_qc_metrics_charge_2}
\end{figure}

Figures \ref{fig:SI_qc_metrics_charge_3} and \ref{fig:SI_qc_metrics_charge_2} summarize the resulting circuits' metrics as a function of active space size for both transpilation approaches. As expected, all of them increase with the number of spin‑orbitals, reflecting the growing complexity of the underlying quantum simulation, or more directly following the LUCJ ansatz scaling. Notably, the largest quantum circuit executed on quantum hardware and used in the SQD post-processing results reported in the main text was run using Q-CTRL Fire Opal. After Fire Opal compression, the circuit had a total depth of 620, a two-qubit-gate depth of 189, 10,592 gates in total, and 1,949 two-qubit gates.

% ========================================
% CHARGE +3
% ========================================

% ========================================
% CHARGE +2
% ========================================

However, across all active space sizes studied, the circuits produced using Q-CTRL's Fire Opal exhibit a systematic and substantial reduction in circuit depth and number of gates relative to the default transpilation, without circuit splitting, approximation or alternative simulation methods as in e.g. entanglement forging\cite{smith2026quantumcentricsimulationhydrogenabstraction}. The gates counts reduction persists across the full range of system sizes considered and becomes increasingly pronounced for larger active spaces, where compilation inefficiencies are typically exacerbated. For the circuit depth, the reduction is more irregular and nuanced, following similar variations in circuits depth yielded by default transpilation as well, yet the depth reductions are still systematic on average, across all experiments.

\begin{figure*}
    \centering
    \includegraphics[width=0.9\linewidth]{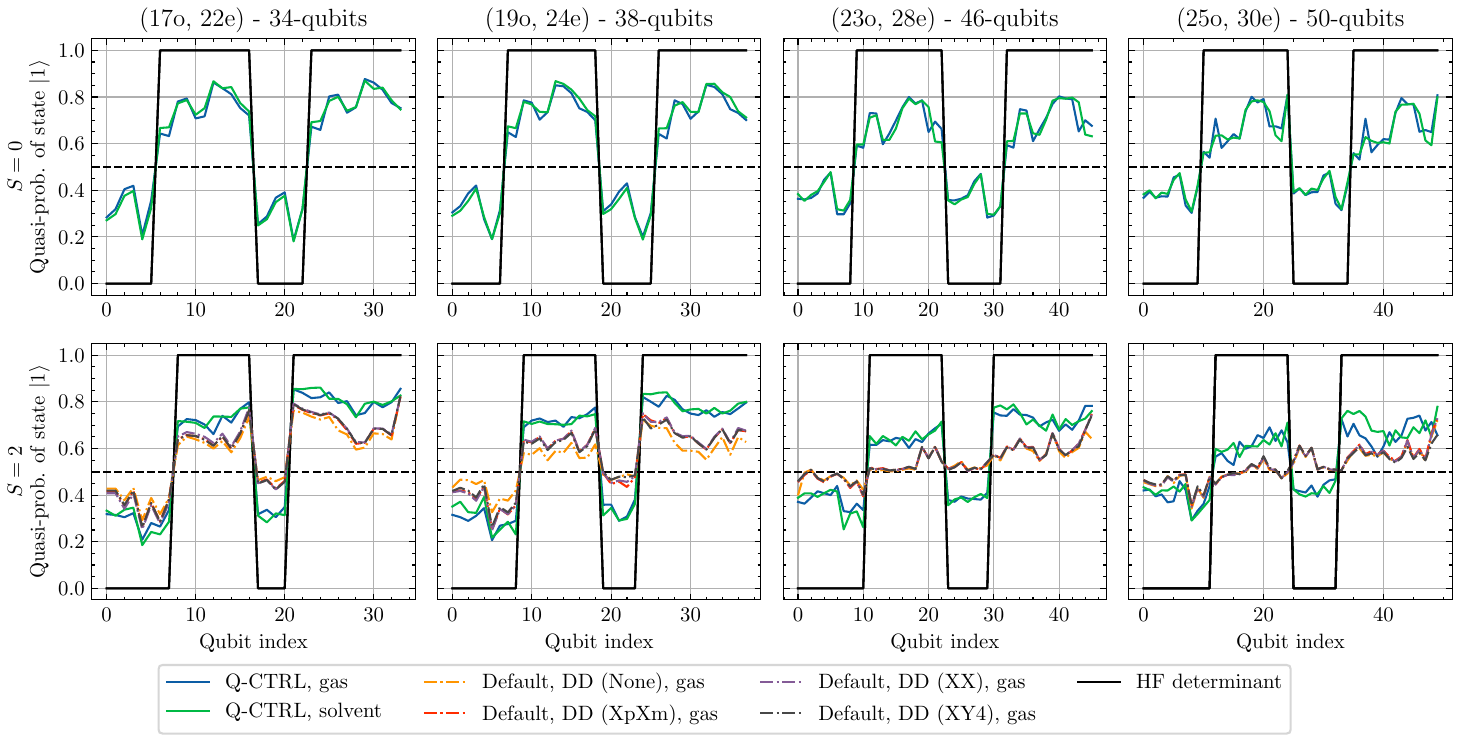}
    \caption{Single-qubit marginal probabilities of measuring $\lvert 1\rangle$ for representative active spaces as a function of qubit index. The upper row shows Q-CTRL's Fire Opal results for the gas-phase and implicit-solvent calculations, while the lower row shows the gas-phase default results obtained without dynamical decoupling and with the $X_pX_m$, $XX$, and $XY4$ dynamical-decoupling sequences. The black line denotes the HF occupation pattern, and the horizontal dashed line at $P(\lvert 1\rangle)=0.5$ denotes the uniformly random single-qubit-marginal reference. Curves represent marginal probabilities averaged across the available molecular geometries. This qubit-resolved visualization is complemented by the per-circuit quantitative analysis in Figure~\ref{fig:SI2_hf_similarity}.}
    \label{fig:SI2_qp_comp}
\end{figure*}

\begin{figure*}
    \centering
    \includegraphics[width=0.95\linewidth]{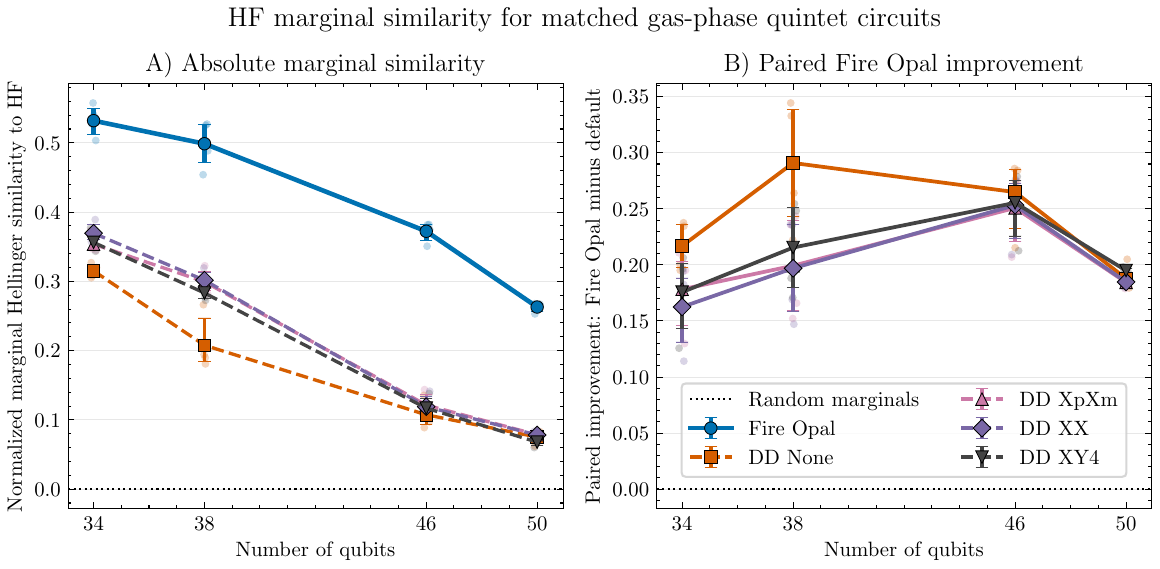}
    \caption{Quantitative comparison of the measured single-qubit occupation marginals with the Hartree-Fock occupation pattern for the charge $+3$ gas-phase quintet calculations. The comparison includes the four molecular geometries for both Q-CTRL's Fire Opal and the default QPU workflows at each active-space size. For every circuit execution, similarity was quantified using the random-normalized marginal Hellinger similarity, for which exact agreement with the HF occupation marginals corresponds to 1 and uniformly random marginals, $P(\lvert 1\rangle)=0.5$ for every qubit, correspond to 0. (A) Absolute marginal similarity for Q-CTRL's Fire Opal and the default workflows without dynamical decoupling (DD) and with the $X_pX_m$, XX, and XY4 dynamical-decoupling sequences. (B) Paired difference between Q-CTRL's Fire Opal and each default workflow for the same active-space size and molecular geometry; positive values indicate greater similarity for Q-CTRL's Fire Opal. Small points represent individual molecular geometries, large markers denote means, and error bars indicate 95\% geometry-level bootstrap confidence intervals. The reported quantity characterizes computational-basis single-qubit marginals and should not be interpreted as quantum-state fidelity.}
    \label{fig:SI2_hf_similarity}
\end{figure*}

Transpiled-circuit metrics provide insight into the practical performance of quantum algorithms executed on hardware, but they constitute only indirect proxies for the quality of the resulting samples. To complement this analysis with a direct examination of the raw QPU results, we evaluated the presence of the Hartree--Fock (HF) determinant in the measured bit-string distributions (Table~\ref{tab:SI1_HF_dets_checkup}) and compared the single-qubit occupation marginals with the corresponding HF occupation pattern (Figures~\ref{fig:SI2_qp_comp} and \ref{fig:SI2_hf_similarity}). For this comparison, the default workflow included executions without dynamical decoupling and with the $X_pX_m$, $XX$, and $XY4$ dynamical-decoupling sequences implemented in Qiskit~\cite{Ezzell2023}. This allowed comparison with Q-CTRL's Fire Opal workflow, which incorporates dynamical decoupling together with additional error-suppression procedures. None of the evaluated approaches introduced sampling overhead in the reported results: all measured shots were retained, such that, for example, \num{500000} shots yielded \num{500000} measured bit strings.

\FloatBarrier

\begin{table}
    \caption{Hartree-Fock determinant presence rates from SQD sampling on \textsf{ibm\_pittsburgh} QPU across active space sizes for a charge 3+ transition metal complex in gas phase and solvent, for singlet and quintet states. Top table shows the QPU runs through whole dissociation pathway consisting 10 data points using Q-CTRL's Fire Opal middleware. Those runs were the input for all the SQD results in the main text. Lower table shows the trial comparison runs for the 4 exemplary data points on 1.44Å, 2.51Å, 3.57Å and 4.63Å dissociation distances. All experiment there were run with gas medium and quintet spin state, with no error suppression or dynamical decoupling in 3 sequence types, as specified in Qiskit. Values ($x/10$, $x/4$) denote the number of successful detections. For each active space size, the number of shots per data point is specified in the second column.}
    \label{tab:SI1_HF_dets_checkup}
    \centering  
    
    A) Q-CTRL's Fire Opal runs
    
    \begin{tabular}{|c|c|c|c|c|c|}
        \hline
        & & \multicolumn{2}{c|}{\textbf{Gas}} & \multicolumn{2}{c|}{\textbf{Solvent}} \\
        \hline
        \textbf{($n_{\textrm{elec}}, n_{\textrm{mo}}$)} & \textbf{Shots} & \textbf{S=0} & \textbf{S=2} & \textbf{S=0} & \textbf{S=2} \\
        \hline
        (22, 17) & \num{200000} & 10/10 & 10/10 & 10/10 & 10/10\\
        \hline
        (24, 19) & \num{300000} & 10/10 & 10/10 & 10/10 & 10/10\\
        \hline
        (28, 23) & \num{500000} & 10/10 & 10/10 & 10/10 & 10/10\\
        \hline
        (30, 25) & \num{500000} & 10/10 & 10/10 & 10/10 & 10/10\\
        \hline
    \end{tabular}
    \vspace{\baselineskip}  
    
    B) Default runs with dynamical decoupling
    
    \begin{tabular}{|c|c|c|c|c|c|}
        \hline
        & & \multicolumn{4}{c|}{\textbf{Gas, S=2}} \\
        \hline
        \textbf{($n_{\textrm{elec}}, n_{\textrm{mo}}$)} & \textbf{Shots} & \textit{None} & \textit{$X_pX_m$} & \textit{XX} & \textit{XY4} \\
        \hline
        (22, 17) & \num{100000} & 4/4 & 4/4 & 4/4 & 4/4\\
        \hline
        (24, 19) & \num{100000} & 3/4 & 4/4 & 4/4 & 4/4\\
        \hline
        (28, 23) & \num{100000} & 0/4 & 0/4 & 0/4 & 0/4\\
        \hline
        (30, 25) & \num{100000} & 0/4 & 0/4 & 0/4 & 0/4\\
        \hline
    \end{tabular}
\end{table}

% ========================================
% Results quality
% ========================================

For the smallest active space, the HF determinant was observed using all execution approaches. At 19 spatial orbitals, one of the four default executions without error suppression did not contain the HF determinant. For the larger active spaces containing 23 and 25 spatial orbitals, corresponding to 46 and 50 qubits, respectively, the HF determinant was not observed in any of the tested default executions, irrespective of the dynamical-decoupling sequence. In contrast, the Q-CTRL's Fire Opal results contained the HF determinant for all dissociation-path geometries, media, spin states, and active-space sizes considered. These observations indicate that dynamical decoupling alone was insufficient to preserve detectable HF support in the tested default executions of the larger circuits.

Figure~\ref{fig:SI2_qp_comp} provides a qubit-resolved view of the measured occupation marginals. For both execution approaches, increasing the active-space size shifts the marginals away from the HF occupation pattern and toward the $P(\lvert 1\rangle)=0.5$ reference associated with uniformly random single-qubit marginals. This trend indicates progressive loss of the occupation information encoded by the LUCJ circuit as circuit complexity increases. Nevertheless, the Q-CTRL's Fire Opal marginals remain visibly closer to the HF occupation pattern than the default results across the active-space sizes considered.

To quantify this behavior, we calculated a random-normalized marginal Hellinger similarity to the HF occupation pattern for each individual circuit execution. For qubit $i$, the measured marginal was represented by the Bernoulli distribution
\begin{equation}
    P_i = (1-p_i,p_i),
\end{equation}
where $p_i=P(X_i=1)$, and the HF reference was represented as
\begin{equation}
    Q_i = (1-h_i,h_i),
\end{equation}
where $h_i\in\{0,1\}$ is the corresponding HF occupation. The mean marginal Hellinger similarity was calculated as
\begin{equation}
    S_{\mathrm{H}} =
    \frac{1}{n}
    \sum_{i=1}^{n}
    \left[
        \sqrt{p_i h_i}
        +
        \sqrt{(1-p_i)(1-h_i)}
    \right],
\end{equation}
and normalized relative to uniformly random marginals according to
\begin{equation}
    S_{\mathrm{H}}^{\mathrm{norm}} =
    \frac{S_{\mathrm{H}}-1/\sqrt{2}}
         {1-1/\sqrt{2}}.
\end{equation}
Under this normalization, exact agreement with the HF occupation marginals corresponds to $S_{\mathrm{H}}^{\mathrm{norm}}=1$, while uniformly random single-qubit marginals correspond to $S_{\mathrm{H}}^{\mathrm{norm}}=0$. This quantity measures similarity of the single-qubit occupation marginals and should not be interpreted as quantum-state fidelity, because it does not capture multi-qubit correlations, relative phases, or similarity to the complete ideal LUCJ state.

The quantitative comparison was restricted to the gas-phase quintet calculations and the four molecular geometries available for both Q-CTRL's Fire Opal and all default workflows, corresponding to dissociation distances of 1.44, 2.51, 3.57, and 4.63~\AA. The similarity was calculated independently for each circuit execution before aggregation. Uncertainty intervals were obtained by bootstrapping the four matched molecular geometries, which were treated as the units of replication; the individual qubit marginals and measurement shots were not treated as statistically independent observations.

As shown in Figure~\ref{fig:SI2_hf_similarity}A, the normalized marginal Hellinger similarity decreased with increasing active-space size for all execution approaches, consistently with progressive noise accumulation in the larger circuits. For Q-CTRL's Fire Opal, the mean similarity decreased from $0.532$ at 34 qubits to $0.263$ at 50 qubits. For the default workflow without dynamical decoupling, it decreased from $0.315$ to $0.075$ over the same range. The default workflows employing dynamical decoupling exhibited the same overall degradation and reached mean similarities between $0.068$ and $0.079$ at 50 qubits.

The paired analysis in Figure~\ref{fig:SI2_hf_similarity}B shows that Q-CTRL's Fire Opal produced higher HF marginal similarity than every default dynamical-decoupling configuration at every active-space size considered. Relative to the default workflow without dynamical decoupling, the mean paired improvement was $0.217$, $0.291$, $0.265$, and $0.188$ at 34, 38, 46, and 50 qubits, respectively. These differences correspond to reductions of approximately $32\%$, $37\%$, $30\%$, and $20\%$, respectively, in the marginal discrepancy from the HF occupation pattern. Comparisons with the $X_pX_m$, $XX$, and $XY4$ dynamical-decoupling workflows yielded consistent results, with discrepancy reductions of approximately $20\%$--$30\%$ across the evaluated active-space sizes. A sensitivity analysis based on normalized mean absolute marginal agreement produced the same qualitative ordering, indicating that the observed advantage was not specific to the Hellinger-based metric.

Overall, the circuit reductions obtained using Q-CTRL's Fire Opal were accompanied by systematically improved raw QPU results. Q-CTRL's Fire Opal produced occupation marginals that remained farther from the random-marginal limit and closer to the HF occupation pattern, while also preserving detectable HF support in the measured bit-string distributions for the largest circuits. The agreement between the Hellinger-based analysis, the mean-absolute-error sensitivity analysis, and the HF-determinant detection rates supports the practical importance of circuit-level optimization and error suppression in near-term SQD workflows.

\FloatBarrier
% ================================================
% Appendix D
% ================================================
\section{Detailed SQD and HCI post-processing info}
\label{app:si_detailed_SQD_and _HCI}

The data reported in Tables~\ref{tab:SI2_SQD_subspaces_charge_3} and \ref{tab:SI2_SQD_subspaces_charge_2} summarize the key hyperparameters controlling the SQD and HCI post-processing workflows. In the SQD procedure, the number of S-CORE iterations was adjusted depending on active-space size and spin state to ensure convergence of both the energy and orbital occupancies, with smaller iteration counts for larger active spaces due to longer iterations runtimes and the total 48hrs limit of the post-processing jobs. In the solvent case, the reported iteration counts reflect the nested SCRF-SQD structure, where a small number of outer SCRF updates is combined with limited inner SQD cycles for efficiency. The resulting variational subspace sizes increase systematically with active-space dimension, reflecting both the exponential growth of Hilbert space and the need for broader sampling coverage, with noticeably larger subspaces required in the gas phase compared to solvent. For HCI, the number of determinants retained in the final ground-state expansion remains relatively stable across gas and solvent conditions, confirming that the chosen thresholds ($\epsilon_1=10^{-3}$, $\epsilon_2=10^{-5}$) provide a balanced and consistent level of correlation treatment across system sizes and spin states.

\begin{table}
    \caption{
    Detailed SQD and HCI post-processing parameters for the [Co(H$_2$O)$_5$CO$_2$]$^{3+}$ system across different active spaces and spin states. (A) Number of S-CORE iterations used in the SQD procedure (for solvent, values are given as $(n_{\mathrm{SCRF}}, n_{\mathrm{inner}})$ corresponding to outer SCRF and inner SQD iterations, respectively). (B) Effective SQD subspace sizes (number of sampled determinants retained in the variational subspace after filtering and batching). (C) Number of determinants in the variational ground-state space obtained from deterministic HCI calculations with fixed thresholds ($\epsilon_1 = 10^{-3}$, $\epsilon_2 = 10^{-5}$, max. num. of iterations $= 26$). All results are reported for charge $+3$ and for both singlet ($S=0$) and quintet ($S=2$) spin states.
    }
    \label{tab:SI2_SQD_subspaces_charge_3}
    \centering
    A) Num. of iterations for SQD
    
    \begin{tabular}{|c|c|c|c|c|}
        \hline
        & \multicolumn{2}{c|}{\textbf{Gas}} & \multicolumn{2}{c|}{\textbf{Solvent}} \\
        \hline
        \textbf{($n_{\textrm{elec}}, n_{\textrm{mo}}$)} & \textbf{S=0} & \textbf{S=2} & \textbf{S=0} & \textbf{S=2} \\
        \hline
        (22, 17) & 50& 50& (6,15)& (6,15)\\
        \hline
        (24, 19) & 50& 50& (6,15)& (6,15)\\
        \hline
        (28, 23) & 11& 17& (6,15)& (6,15)\\
        \hline
        (30, 25) & 7& 13& (5,8)& (6,15)\\
        \hline
    \end{tabular}
    \vspace{\baselineskip}  
    
    B) Subspace sizes for SQD  ($\times 10^6$)
    
    \begin{tabular}{|c|c|c|c|c|}
        \hline
        & \multicolumn{2}{c|}{\textbf{Gas}} & \multicolumn{2}{c|}{\textbf{Solvent}} \\
        \hline
        \textbf{($n_{\textrm{elec}}, n_{\textrm{mo}}$)} & \textbf{S=0} & \textbf{S=2} & \textbf{S=0} & \textbf{S=2} \\
        \hline
        (22, 17) & 7 & 2.5 & 4 & 1 \\
        \hline
        (24, 19) & 12 & 5.5& 5& 2\\
        \hline
        (28, 23) & 22& 10& 12& 5\\
        \hline
        (30, 25) & 27& 17& 13& 6\\
        \hline
    \end{tabular}
    \vspace{\baselineskip}  

    C) Groundstate determinants for HCI ($\times 10^3$)
    
    \begin{tabular}{|c|c|c|c|c|}
        \hline
        & \multicolumn{2}{c|}{\textbf{Gas}} & \multicolumn{2}{c|}{\textbf{Solvent}} \\
        \hline
        \textbf{($n_{\textrm{elec}}, n_{\textrm{mo}}$)} & \textbf{S=0} & \textbf{S=2} & \textbf{S=0} & \textbf{S=2} \\
        \hline
        (22, 17) & 3.6& 5.3& 3.6& 3.5 \\
        \hline
        (24, 19) & 4.0& 6.0& 3.9& 4.3 \\
        \hline
        (28, 23) & 6.6& 9.0& 6.7 & 7.6 \\
        \hline
        (30, 25) & 8.3& 11.6& 8.3& 10.5 \\
        \hline
    \end{tabular}
\end{table}

\begin{table}
    \caption{
    Detailed SQD and HCI post-processing parameters for the [Co(H$_2$O)$_5$CO$_2$]$^{2+}$ system across different active spaces and spin states. (A) Number of S-CORE iterations used in the SQD procedure (for solvent, values are given as $(n_{\mathrm{SCRF}}, n_{\mathrm{inner}})$ corresponding to outer SCRF and inner SQD iterations, respectively). (B) Effective SQD subspace sizes (number of sampled determinants retained in the variational subspace after filtering and batching). (C) Number of determinants in the variational ground-state space obtained from deterministic HCI calculations with fixed thresholds ($\epsilon_1 = 10^{-3}$, $\epsilon_2 = 10^{-5}$, max. num. of iterations $= 26$). All results are reported for charge $+2$ and for both doublet ($S=0.5$) and quartet ($S=1.5$) spin states.
    }
    \label{tab:SI2_SQD_subspaces_charge_2}
    \centering  
    A) Num. of iterations for SQD
    
    \begin{tabular}{|c|c|c|c|c|}
        \hline
        & \multicolumn{2}{c|}{\textbf{Gas}} & \multicolumn{2}{c|}{\textbf{Solvent}} \\
        \hline
        \textbf{($n_{\textrm{elec}}, n_{\textrm{mo}}$)} & \textbf{S=1/2}& \textbf{S=3/2}& \textbf{S=1/2}& \textbf{S=3/2}\\
        \hline
        (23, 17) & 50& 50& (5,10)& (5,10)\\
        \hline
    \end{tabular}
    \vspace{\baselineskip}  
    
    B) Subspace sizes for SQD  ($\times 10^6$)
    
    \begin{tabular}{|c|c|c|c|c|}
        \hline
        & \multicolumn{2}{c|}{\textbf{Gas}} & \multicolumn{2}{c|}{\textbf{Solvent}} \\
        \hline
        \textbf{($n_{\textrm{elec}}, n_{\textrm{mo}}$)} & \textbf{S=1/2}& \textbf{S=3/2}& \textbf{S=1/2}& \textbf{S=3/2}\\
        \hline
        (23, 17) & 1.1& 0.9& 0.8& 0.5\\
        \hline
    \end{tabular}
    \vspace{\baselineskip}  

    C) Groundstate determinants for HCI ($\times 10^3$)
    
    \begin{tabular}{|c|c|c|c|c|}
        \hline
        & \multicolumn{2}{c|}{\textbf{Gas}} & \multicolumn{2}{c|}{\textbf{Solvent}} \\
        \hline
        \textbf{($n_{\textrm{elec}}, n_{\textrm{mo}}$)} & \textbf{S=1/2}& \textbf{S=3/2}& \textbf{S=1/2}& \textbf{S=3/2}\\
        \hline
        (23, 17) & 3.5& 2.8& 3.7& 2.6\\
        \hline
    \end{tabular}
\end{table}

\FloatBarrier

\newpage
\bibliography{jctc_bib.bib}